\documentstyle[epsfig]{kapproc} 

\kluwerbib

\startauthorindex

\begin{document}

\articletitle{A Method to Search for Associations of Young Stars
\footnote{
Based on observations made under the
ON-ESO
agreement for the joint operation of
the 1.52\,m ESO telescope and at the  Observat\'{o}rio do Pico dos Dias
(LNA/MCT),  Brazil}
}
\author{Germano R. Quast, Carlos A. O. Torres}
\affil{Laborat\'{o}rio Nacional de Astrof\'{\i}sica/MCT,
 37504-364 Itajub\'{a}, Brazil}
\email{germano@lna.br, beto@lna.br}
\author{Claudio H. F. Melo, Michael Sterzik}
\affil{ESO,  
Cassilla 19001 Santiago 19, Chile}
\email{cmelo@eso.org, msterzik@eso.org}
\author{Ramiro de la Reza,  Licio da Silva}
\affil{Observat\'{o}rio Nacional/MCT,
20921-030 Rio de Janeiro, Brazil}
\email{delareza@on.br, ldasilva@eso.org}

\chaptitlerunninghead{SACY}

\anxx{Torres et al.}

\begin{abstract}
In the SACY 
(Search for Associations Containing Young-stars) 
project we try to identify associations
of stars younger than the Local Association
among HIPPARCOS and/or TYCHO-2 stars later than G0
which are counterparts of the
ROSAT X-ray bright sources.
High-resolution spectra for the possible optical counterparts
were obtained in order to assess both the
youth and the spatial motion of each target.
More than 1000 ROSAT sources were observed, covering a large
area in the Southern Hemisphere.
Associations are characterized mainly by the similarity in UVW velocity 
space of their proposed member, but other parameters, as
evolutionary age, Li abundance and distribution in space must
also be taken into account.
We proposed a method to identify associations when proper motions
and radial velocities are available, but no parallaxes.
Using the method we found eleven associations in the SACY data.

\end{abstract}


\section*{Introduction}
In 1989, de la Reza et al. searched for isolated T Tau stars (TTS)
and found a group of TTS around TW Hya.
This was the beginning of the Pico dos Dias survey (PDS). 
The PDS was a search for young stars using the IRAS Point
Source Catalog as the main selector (Gregorio-Hetem et al., 1992;
Torres et al., 1995; Torres, 1998).
X-ray sources  from the ROSAT All-Sky Survey (RASS)  gave a new
tool to find new young associations (Neuh\"{a}user 1997).
With some of these sources, Torres et al. (2000) found evidences for a
young nearby association: they  called it Horologium Association
(HorA). Almost simultaneously, Zuckerman \& Webb (2000) found
another one, very similar and adjacent in the sky, which they
called Tucana Association (TucA). In order to examine the physical
relation between both of them, and to search for other ones, we
started the SACY project (de la Reza et al. 2001; Torres et al.
2001; Quast et al. 2001).

\section{Observations}

For SACY we selected all bright RASS sources that
could be associated with
TYCHO-2 or HIPPARCOS stars with (B-V) $>$ 0.6, 
excluding well known
RS CVn, W UMa, giants, etc from  SIMBAD. 
We restricted our sample to stars later than G0
because we use the Li I 6707\AA\ equivalent width as
an age indicator.

We obtained high resolution spectra for the selected candidates with
the FEROS \'echelle spectrograph (Kaufer et al. 1999)
(resolution of 50000; spectral coverage of 5000\,\AA)
of the 1.52 m ESO telescope at La Silla
or  with the coud\'{e} spectrograph
(resolution of 9000; spectral coverage of 450\,\AA, centered at 6500\,\AA)
of  the 1.60\,m telescope of the Observat\'{o}rio do Pico dos Dias (OPD).
For some stars we obtained
radial velocities with  CORALIE at the Swiss Euler Telescope at ESO
(Queloz et al. 2000). We derived spectral classifications,
radial velocities and equivalent widths of {Li\,I 6707\,\AA} lines.
In particular, the Li\,I line is important since it can provide
a crude age estimate (Jeffries 1995) for late type stars.
If the Li\,I line equivalent width is larger than the
highest values for stars stars belonging to the Local
Association  (Neuh\"{a}user 1997), the star is  flagged as young.

\mbox{$UBV(RI)_C$} photometry for some of the stars of the sample  
was obtained   using FOTRAP (Jablonski et al. 1994) 
at  the 0.60\,m   telescope of the  OPD.

\subsection{Statistics}

There are 9574 RASS bright sources in the Southern Hemisphere,
2071 of them having counterparts with (B-V) $>$ 0.6 in TYCHO-2.
We observed 1096 sources. We also used published information for
99 others, most of them without interest for our search and
the young ones taken mainly from Covino et al. (1997).  
Unfortunately  we have no observations in the Upper Cen-Lup 
of the Sco-Cen Association.
We classified 201 stars as giants and 966 as dwarfs,
421 of them being younger than the Pleiades,
174 having the Pleiades age and 371 older than it.

More details can be seen in the paper of Torres et al.
in the meeting book. 
 
\section{The analysis}

Our observational data set include proper motions, radial velocities, 
spectral types, B-V magnitudes, Li abundances and, for some stars, 
HIPPARCOS parallaxes.
For B-V colors we use data with the following 
priorities: our own colors observations, good observed colors 
found in the literature, HIPPARCOS colors, TYCHO-2 colors 
corrected to the Johnson system if the stars are brighter than 
magnitude 10 and finally, intrinsic colors related to their
spectral type.

As most of our young stars have no HIPPARCOS parallaxes we developed
a method to find possible associations:
We first perform a global kinematical analysis in which, for each point 
in a grid in UVW space, we compute the density of stars around this 
point, adjusting the parallaxes so that the distances in UVW space to 
this point (or the moduli of the space velocity vectors relative
to this point) are minimized. 
Since we  preserve the parallaxes of HIPPARCOS stars their
UVW are fixed.
Constrains are applied so that the stars fall 
into plausible positions in the HR diagram. 
Maxima density concentrations much larger than
the background fluctuation may 
represent real associations. 

Once a possible association has been found by the global analysis 
and its age  estimated (i. e., by the Li abundance), 
its parameters can be refined and the membership to the association
for each star weighted, 
 by a simultaneous adjustment of the 
kinematics and the age.
We impose some restriction on the magnitudes differences
that we accept possible candidates. 
For example, we can reject a star if its kinematical
solution for a possible membership to an association
age is more than one magnitude. 
We weight the ``evolutionary distance" and the ``kinematical distance"
to each star solution to accept or reject it as a possible member. 
Since the kinematics of the young associations are, in general,
very similar one to another we have to use 
the distribution in space  as an additional constrain. 
In fact, all the main concentrations in UVW space 
are also spatially localized, albeit
some of them cover large areas.
It is in general very easy to recognize that an initial concentration
in UVW is actually formed by more than one association with
similar velocities as its split in distinct concentrations
in the real space. 
Anyway, for some stars we must infer a ``best  membership".
Some iterations may be necessary to obtain a good solution. 

\section{The Young Associations}

Applying the method to the database of the SACY project
we find 11 associations of young stars.
Their relevant properties are shown in the Tables 1 and 2
and could be visualized in the figures. For each
suggested association we plot:

a) The distribution in the sky in a polar projection.

b) The distribution of {Li\,I 6707\,\AA}  line equivalent width
as a function of color. 
The continuous line represents the upper limit for Pleiades Association stars.

c) The evolutionary diagram. The continuous lines represent the
ZAMS and an isochrone of either 30 Myr or 10 Myr from Siess (1997)
\begin{table}[ht]
\caption
{Space motions and parallaxes of the Young Associations}
\begin{tabular}{lrrrrrrrrr}
\sphline

Name&U&$\sigma$&V&$\sigma$&W&$\sigma$&$\pi$&$\sigma$&N$_*$\\
&km/s&km/s&km/s&km/s&km/s&km/s&mas&mas\\
\sphline
GAYA1&-9.1&1.1&-20.9&1.0&-1.2&0.9&22.0&2.2&16\\
GAYA2&-11.0&1.0&-22.5&1.0&-4.6&1.1&11.9&4.2&41\\
TWA&-12.1&0.7&-17.1&0.8&-5.5&0.9&22.4&5.6&5+3\\
$\epsilon$ ChaA&-7.7&0.6&-19.7&0.8&-8.5&0.9&10.5&1.3&15\\
LCC&-8.6&0.8&-21.4&1.1&-5.6&1.1&8.7&1.4&40\\
US&-4.7&1.2&-19.3&0.7&-5.2&1.3&7.5&1.4&43\\
YSSA&-4.0&1.3&-13.4&1.0&-8.0&1.2&8.6&1.9&21+5\\
$\beta$ PicA&-9.5&1.2&-16.4&1.1&-9.4&0.9&34.1&27.2&16+1\\
OctA&-10.4&0.6&-1.5&0.6&-8.0&0.8&8.9&1.5&6\\
ArgusA&-21.5&0.8&-13.1&1.1&-5.1&1.4&9.8&2.7&14\\
AnA&-7.1&0.8&-28.0&1.1&-12.4&1.2&19.0&10.5&11\\
\sphline

\end{tabular}

\end{table}
\begin{table} [ht]
\caption{Positions relative to the Sun and ages}
\begin{tabular}{lrrrrrrrrr}
\sphline

Name&X&$\sigma$&Y&$\sigma$&Z&$\sigma$&$\rho_{max}$&age&N$_*$\\
&pc&pc&pc&pc&pc&pc&pc&Myr&\\
\sphline

GAYA1&12&14&-25&8&-33&5&26&30&16\\
GAYA2&7&27&-78&33&-31&25&84&20&41\\
TWA&9&6&-41&10&20&5&18&8&8\\
$\epsilon$ ChaA&47&7&-82&12&-6&14&34&10&15\\
LCC&57&12&-100&20&16&17&75&15&40\\
US&132&27&-28&18&22&18&72&8&43\\
YSSA&117&22&-10&10&-2&30&50&8&26\\
$\beta$ PicA&33&28&-8&16&-16&7&50&15&17\\
OctA&65&30&-73&7&-55&6&46&30&6\\
ArgusA&30&30&-95&34&-10&25&83&30?&14\\
AnA&4&32&-35&28&-35&17&69&50&11\\
\sphline

\end{tabular}
\end{table}

In Tables 1  \& 2 we present the properties of the  young
associations detected in this way. In Table 1 we give the mean
kinematical values and the mean parallax.
For some known associations we use
$bonafide$ members not observed in the SACY to help in their definition.
Their numbers are
indicated in  the last column in Table 1.
In Table 2 we give the mean XYZ, the age, 
and the distance ($\rho_{max}$)
of the most distant member with respect to the calculated center of
the association, giving an idea of its size.

\subsubsection{GAYA1} 

We are calling GAYA (Torres et al. 2001)
two nearby concentrations on the UVW space,
separated mainly in W velocities. Both seem adjacent in the real
space. GAYA1 is somewhat older and is one of the more well 
defined of the associations
in SACY and the previous HorA and TucA are within it.  
Some of the proposed members  of TucA are outside of
the velocity definition (mainly the eastern ones).
Of their 16 proposed members 8 have parallaxes in HIPPARCOS.
The spread in distance is small and this does not seem an artifact either
of the SACY or our analysis as its derived center is at only 45\,pc. 

\subsubsection{GAYA2}

GAYA2 is much  less well defined, although it shows a clear concentration
using HIPPARCOS stars, but reinforced by members of Lower Cru-Cen (LCC). 
Actually, the UVW are very near the LCC ones 
and it is adjacent in space.
Nevertheless GAYA2 seems older than LCC and closer to us.

\subsubsection{TWA} 

The TW Hya association is not very well defined in SACY
since only two members have trigonometric parallaxes. 
Torres et al. (2003) present a list of proposed members, but many of them lack
information for a complete kinematical analysis. 
Anyway, we try to use all possible data. 
The convergence method has problems as TWA is near in velocity and
space of $\epsilon$ ChaA and LCC. 
We applied it limiting the possible spatial volume  but
including any star position in Torres et al. list. 
Nevertheless our solution excludes many stars in their list. 
We proposed as {\it bonafide} kinematical members: TWA 1, 2, 3, 4, 7, 8, 12
plus a new member, CD-39\,7538.
\subsubsection{$\epsilon$ ChaA}

This association is defined by 
Mamajek, Lawson \& Feigelson (2000).
We propose new members, enlarging it.
There is ambiguity for about half of the proposed members
between $\epsilon$ ChaA and LCC, but the solutions show a
consistent separation in UVW space.
One of the proposed members is PDS66.
The distance found by us indicates it is in front of the Cha complex.
\subsubsection{LCC} 

This association has UVW near those of  $\epsilon$\,ChaA
and GAYA2 and the age seems between both. 
The LCC found in SACY is very similar to that found by Sartori et al. (2003)
for early-type stars. In our solution TWA19 belongs to LCC.
\subsubsection{US} 

The Upper Sco (US) has UVW near those of LCC and YSSA. 
US can easily be separated from LCC in real space, but many stars 
may be assigned both to US and YSSA. 
Since we have almost no observations in Upper Cen-Lupus
we can not say if they would be separated in SACY.    
\subsubsection{YSSA}

This is a group of young stars, spread from $\rho$ Oph to R Cra,
with very similar properties, that we are now calling the
Young Sco-Sgr Association.
The western border engulfs the stars mentioned in  
Quast et al. (2001) and   Neuh\"{a}user  et al. (2000).
The split in space distribution can be explained by
the incompleteness in the RASS coverage.
Anyway, the convergence process gives some superposition with US association.
The distance is near the assumed one for the R CrA cloud.
\subsubsection{$\beta$ PicA} 

As described by Zuckerman et al.(2001), this association
is very close to the Sun.
We propose new members, some of then as far as 80\,pc, but the
distribution in space seems very consistent.
Among the new proposed members is V4046 Sgr, a notorious object,
classified before as an isolated SB classical TTS
(de la Reza et al., 1986; Quast, 1998; Quast et al. 2000).
WW PsA and TX PsA could be members (Song et al. 2002), 
but their parallaxes should be 49.5\,mas, closer than HIPPARCOS ones, 
about 2$\sigma$ of the HIPPARCOS errors.
\subsubsection{OctA} 
 
This is a very homogeneous small group of almost aligned stars
(all young G stars)  near the South Celestial Pole.
Since this region belongs to a completely surveyed area of the SACY,
new members have to be found by other means.
\subsubsection{ArgusA} 

Although not very well defined, it
has a special position in UVW. Since many stars are in Car, Vel and Pup
we tentatively propose to call it as Argus A.
\subsubsection{AnA} 

Like ArgusA, the main reason to claim
for this possible association are the very special UVW.
The majority of the proposed members 
have parallaxes and, therefore, this is one of the concentrations
in the HIPPARCOS sample.

GAYA1, GAYA2, LCC, US and YSSA form a decreasing sequence 
in age, going from west to east, 
and they seem to form a kind of continuum in UVW space.
All the associations but the last three in the 
tables can represent local aspects of a global star forming process.

\begin{acknowledgments}
This work was partially supported by a CNPq - Brazil grant to C. A. O. Torres
(pr. 200356/02-0).

\end{acknowledgments}

\pagebreak
\begin{figure}[ht]
\centerline{\psfig{file=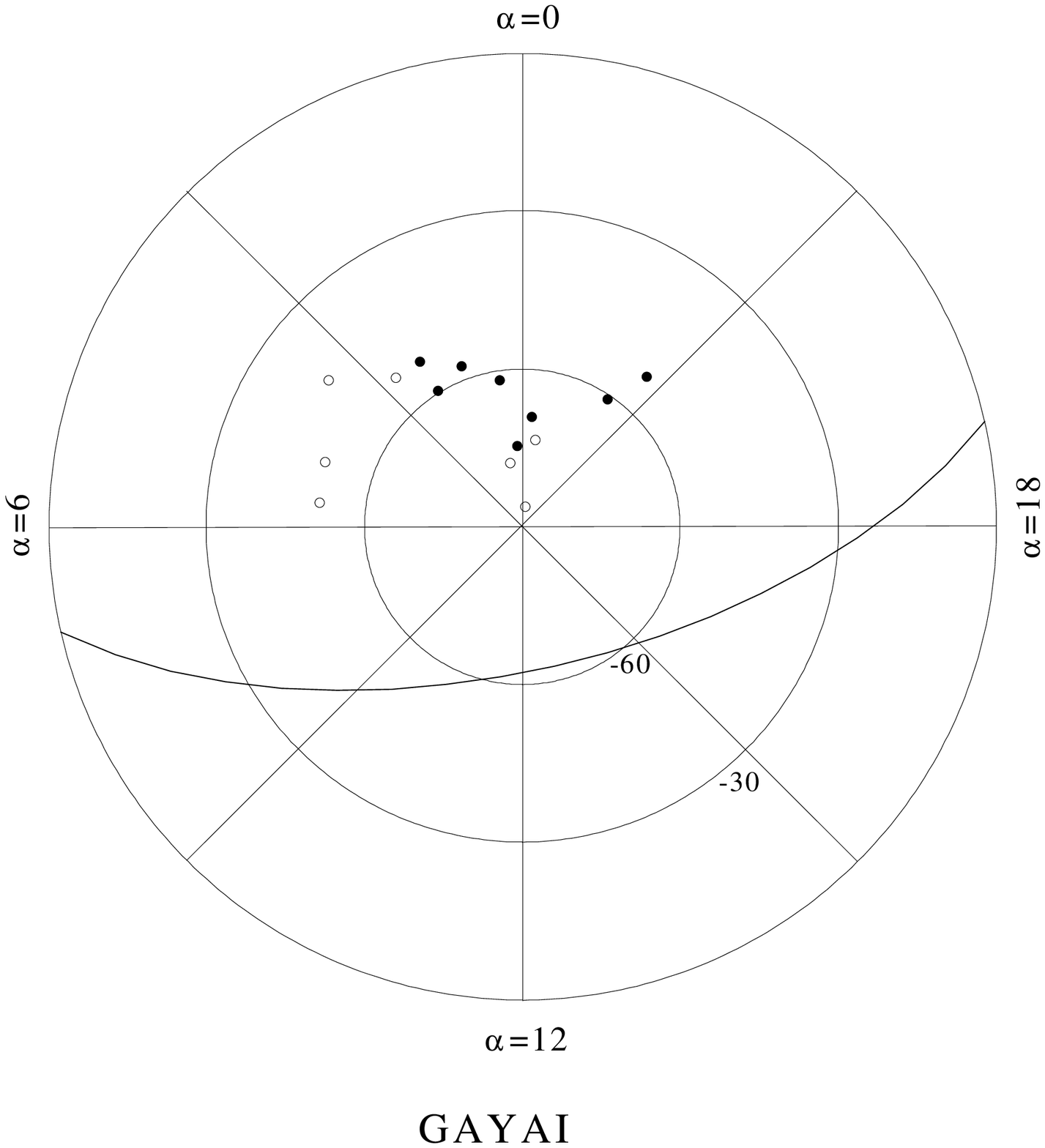,height=8cm}}
\caption{Polar representation of the GAYA1 stars, where the transversal
line is the Galatic Equator. Here and in the next figures, black points
represent HIPPARCOS stars}
\end{figure} 

\begin{figure}[ht]
\sidebyside
{{\psfig{file=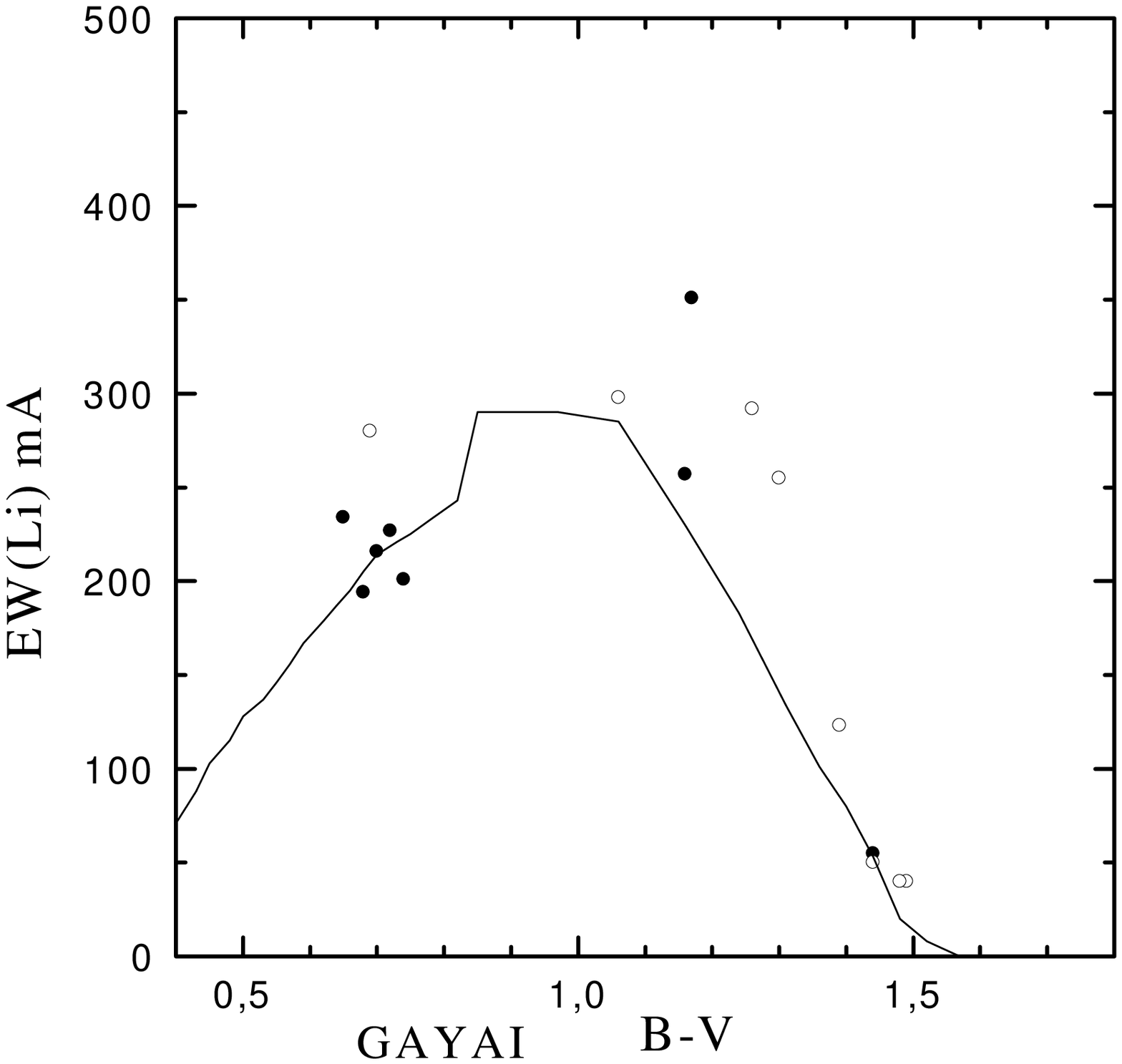,width=5.7cm}}
\caption{Distribution of Li line equivalent width for GAYA1.}}
{{\psfig{file=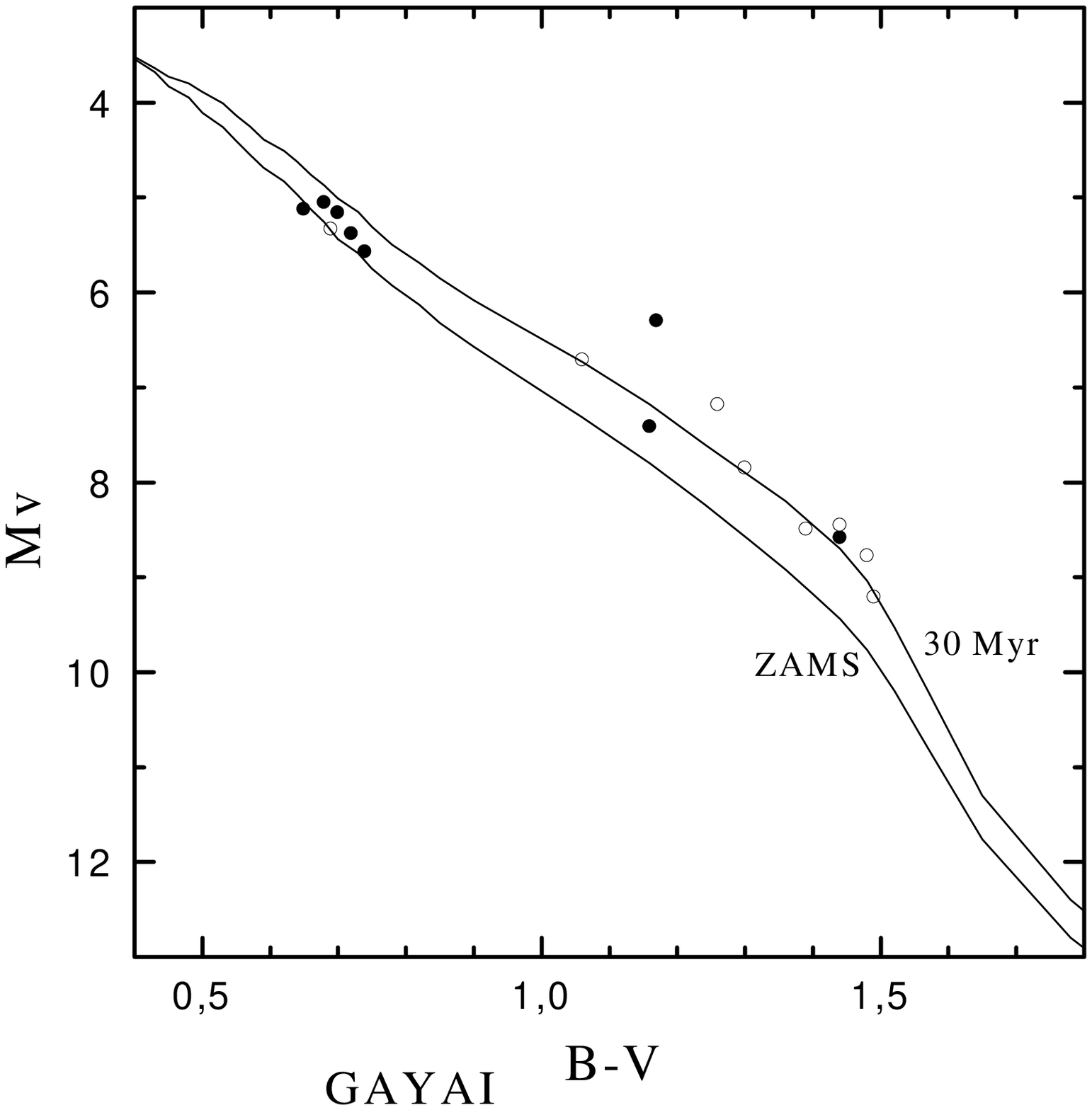,width=5.7cm}}
\caption{Evolutionary diagram for GAYA1.}}
\end{figure}
\pagebreak
\begin{figure}[ht]
\centerline{\psfig{file=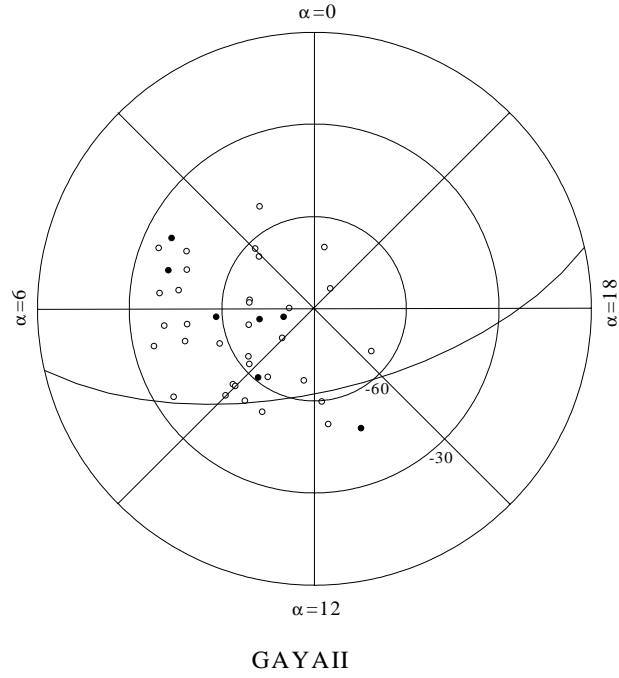,height=9cm}}
\caption{Polar representation of the GAYA2 stars.}
\end{figure} 

\begin{figure}[ht]
\sidebyside
{{\psfig{file=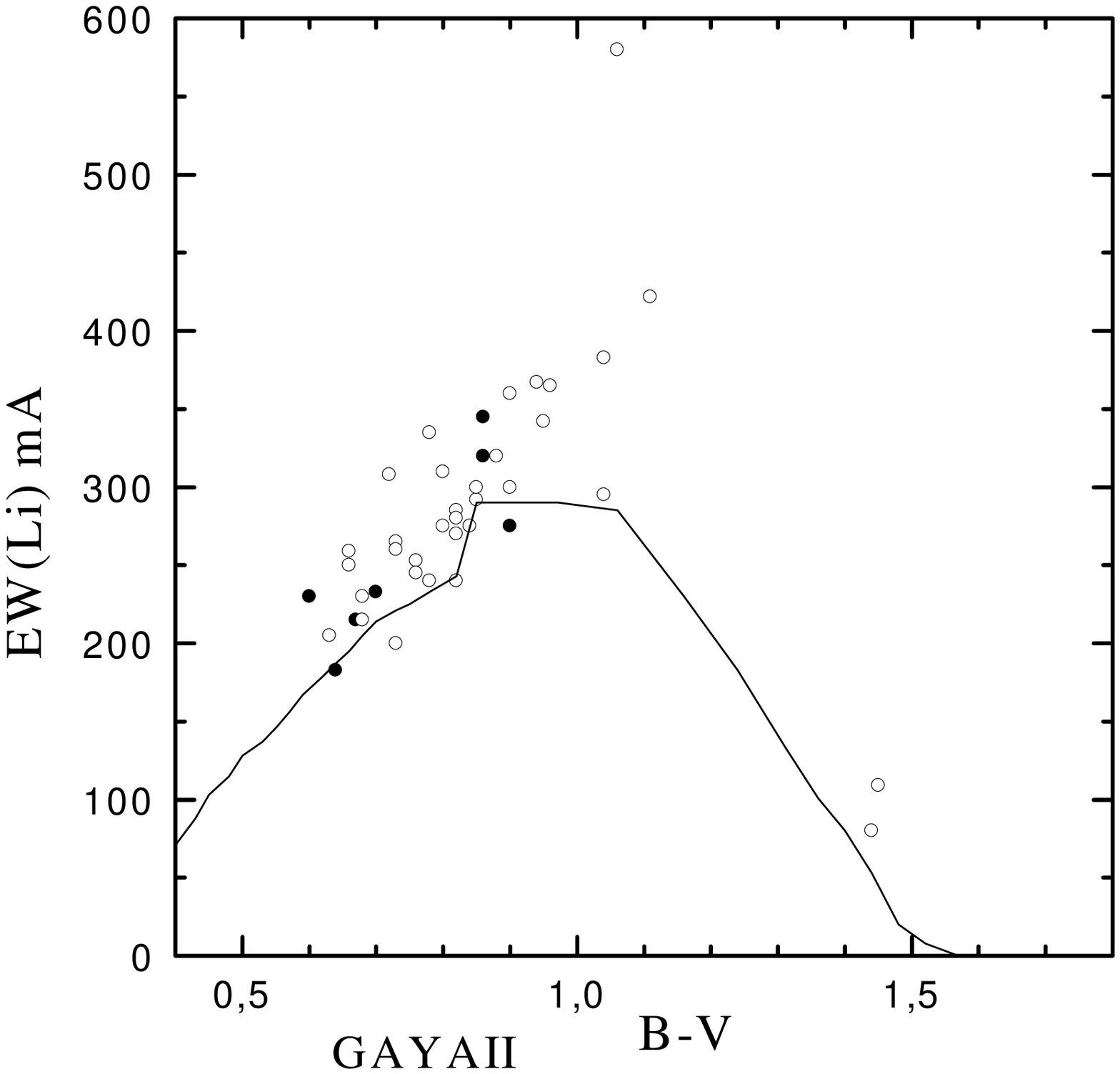,width=5.7cm}}
\caption{Distribution of Li line equivalent width for GAYA2.}}
{{\psfig{file=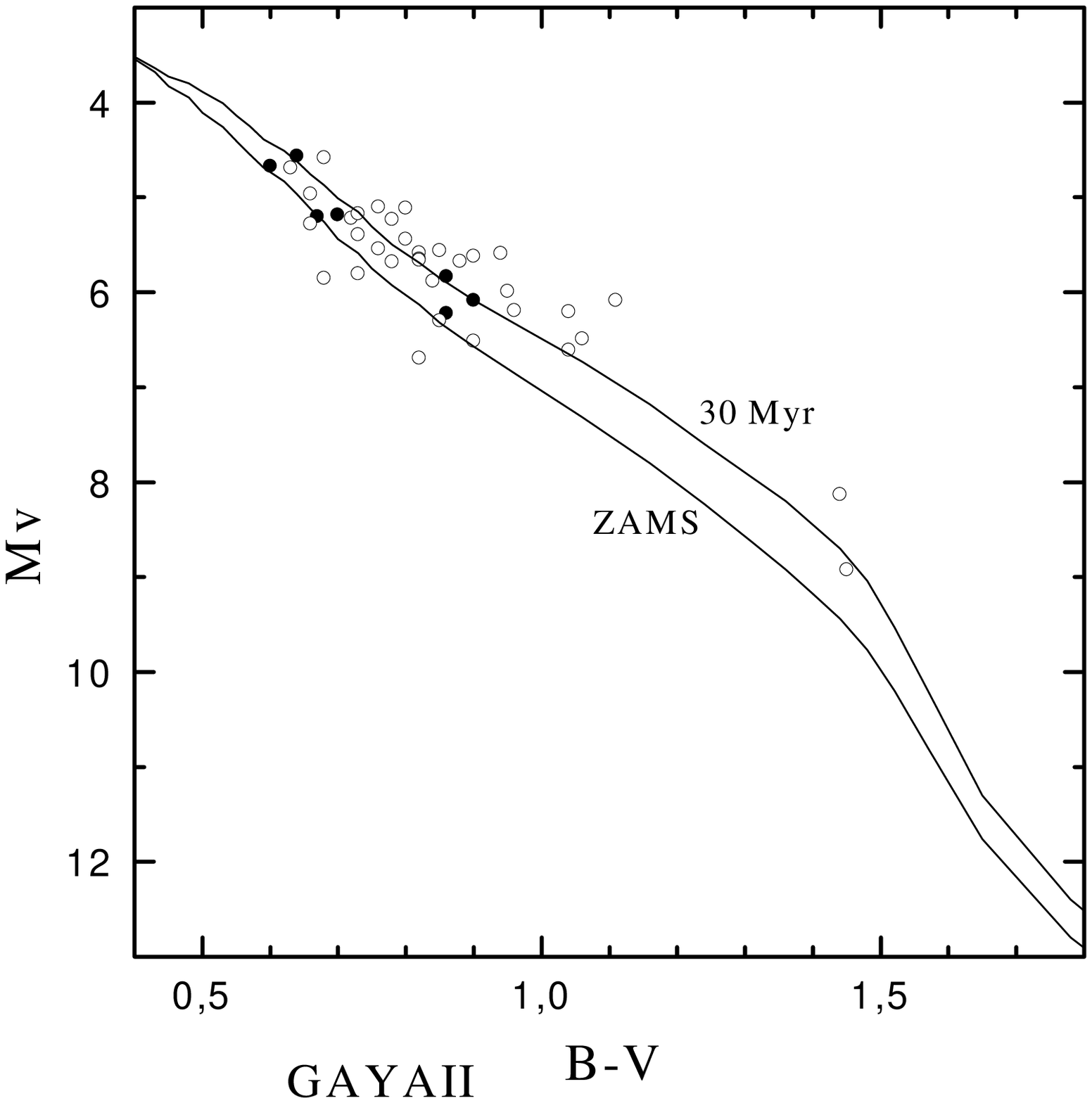,width=5.7cm}}
\caption{Evolutionary diagram for GAYA2.}}
\end{figure}
\pagebreak
\begin{figure}[ht]
\centerline{\psfig{file=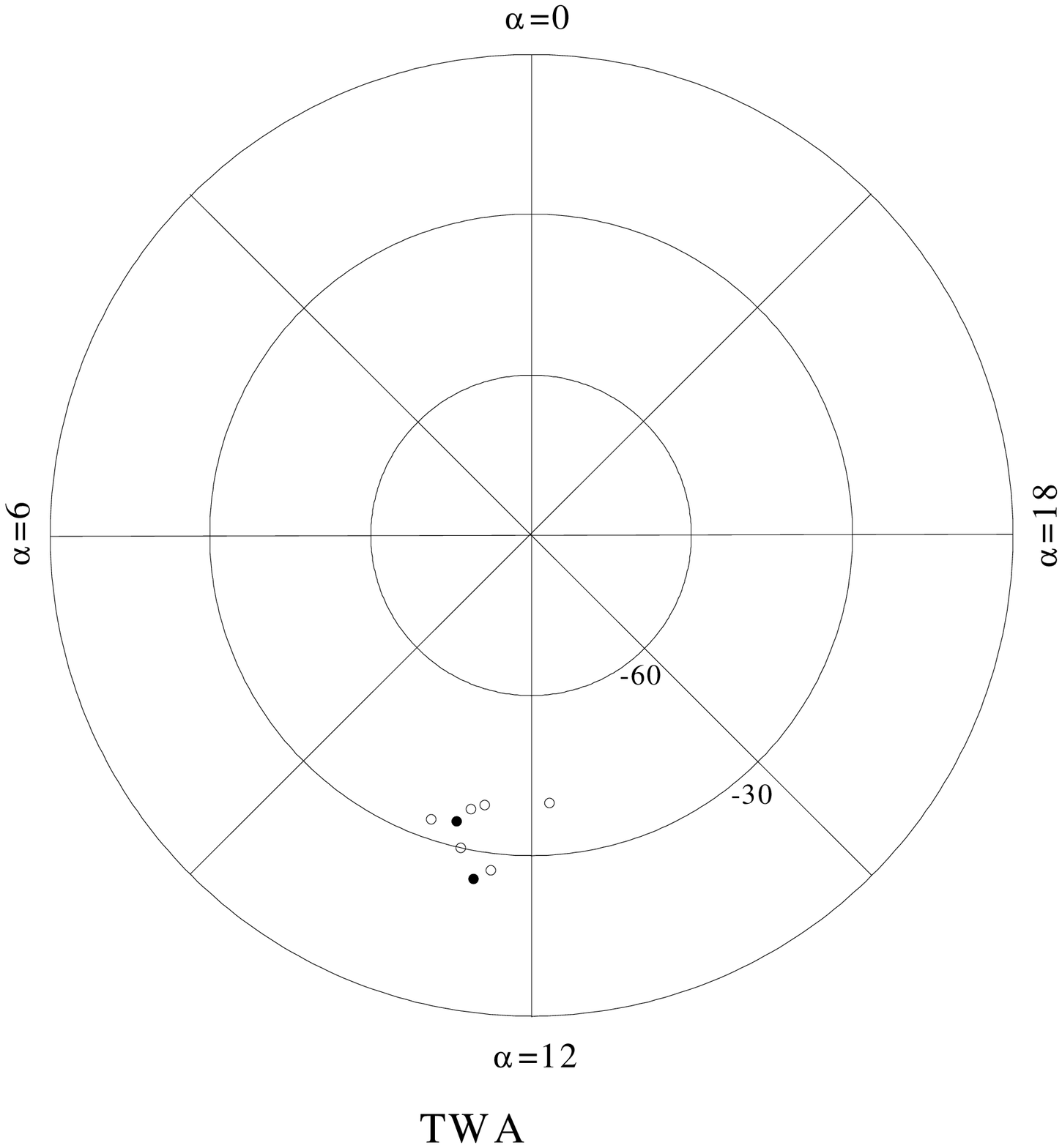,height=9cm}}
\caption{Polar representation of the TWA stars.}
\end{figure} 

\begin{figure}[ht]
\sidebyside
{{\psfig{file=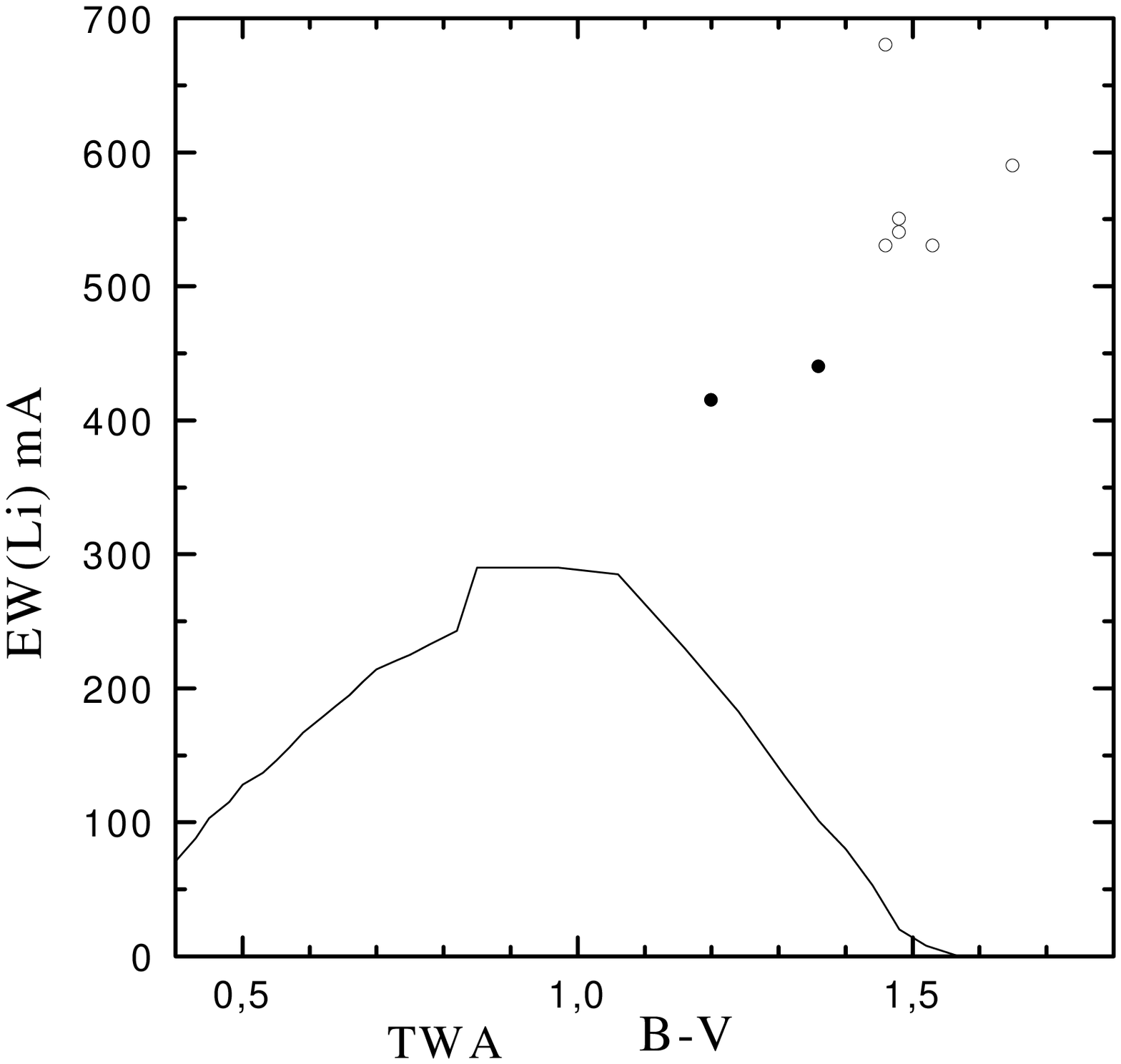,width=5.7cm}}
\caption{Distribution of Li line equivalent width for TWA.}}
{{\psfig{file=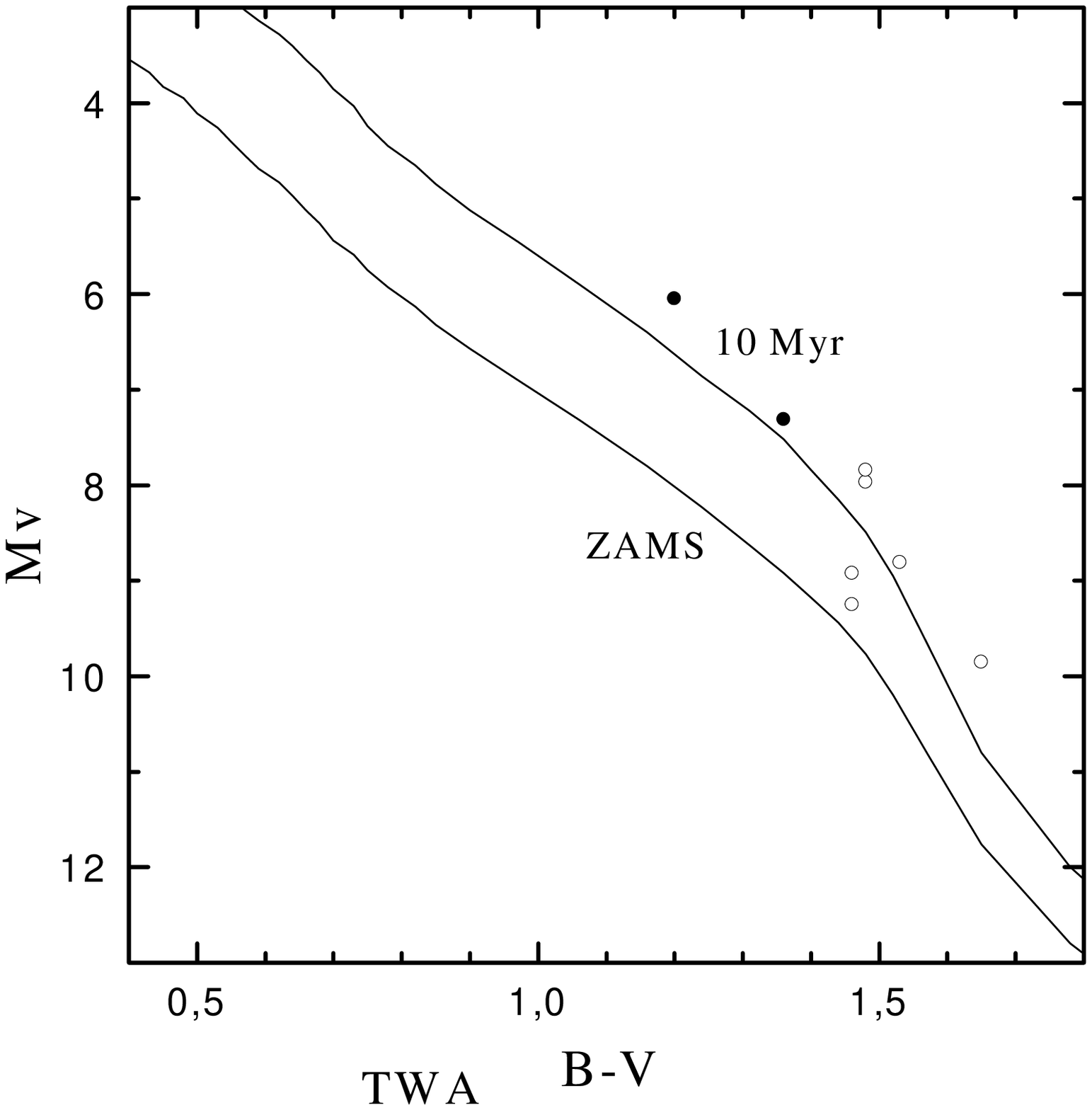,width=5.7cm}}
\caption{Evolutionary diagram for TWA.}}
\end{figure}
\pagebreak
\begin{figure}[ht]
\centerline{\psfig{file=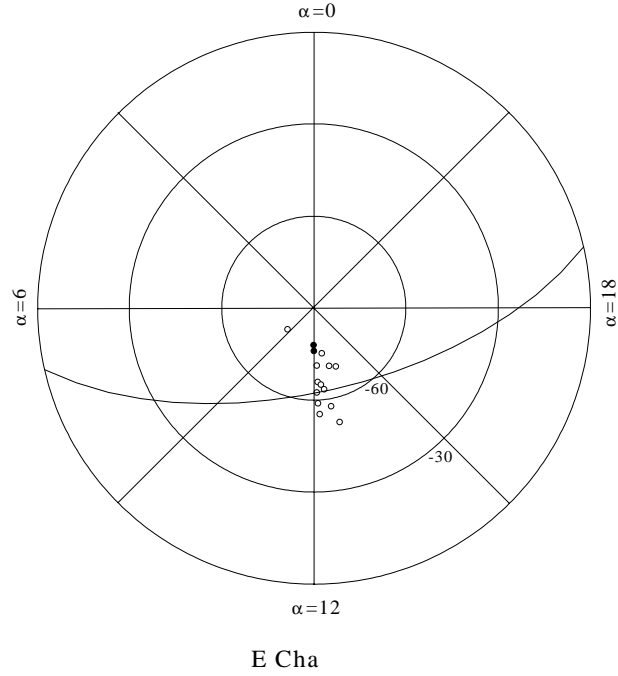,height=9cm}}
\caption{Polar representation of the $\epsilon$ ChaA  stars.}
\end{figure} 

\begin{figure}[ht]
\sidebyside
{{\psfig{file=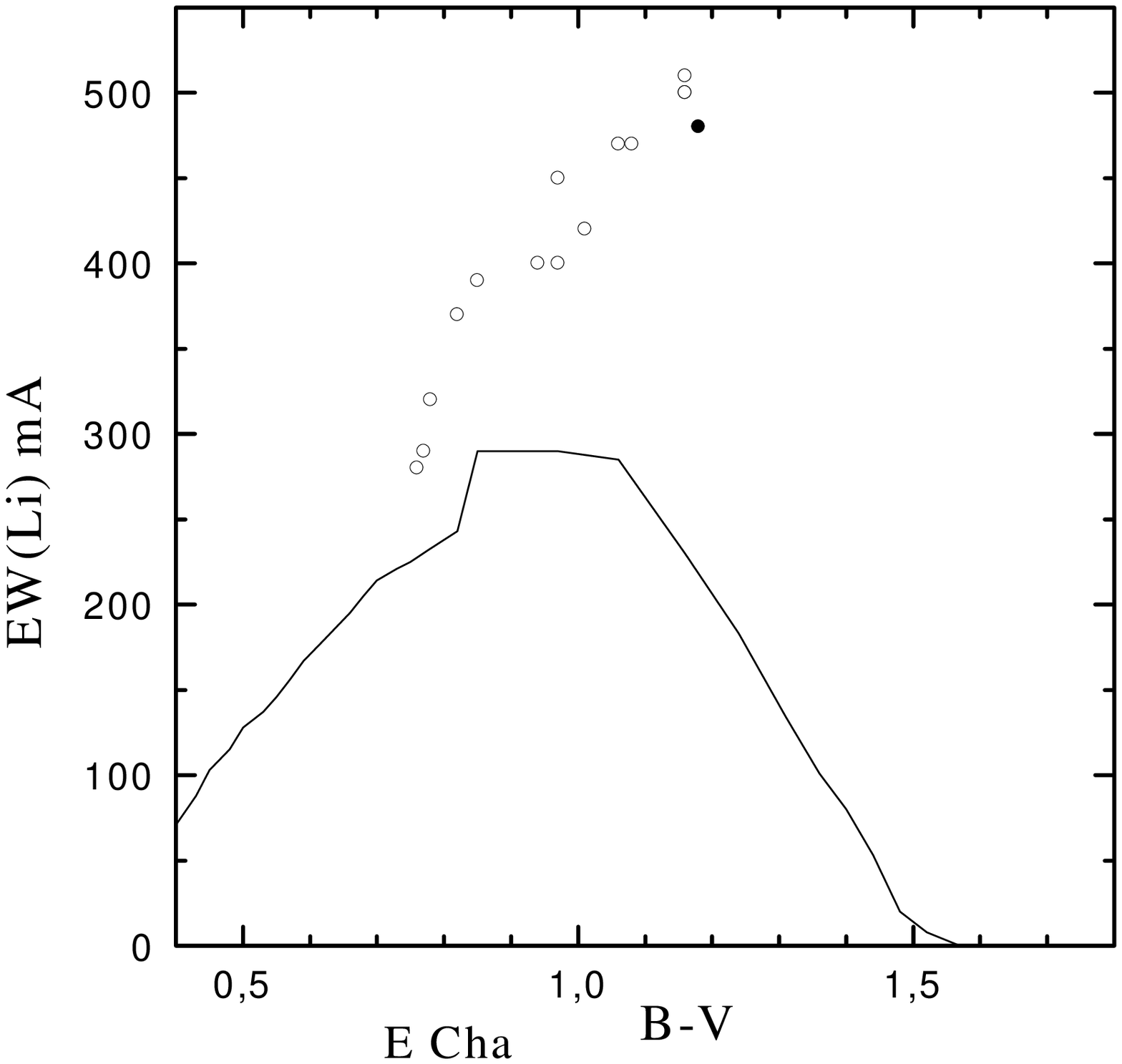,width=5.7cm}}
\caption{Distribution of Li line equivalent width for $\epsilon$ ChaA .}}
{{\psfig{file=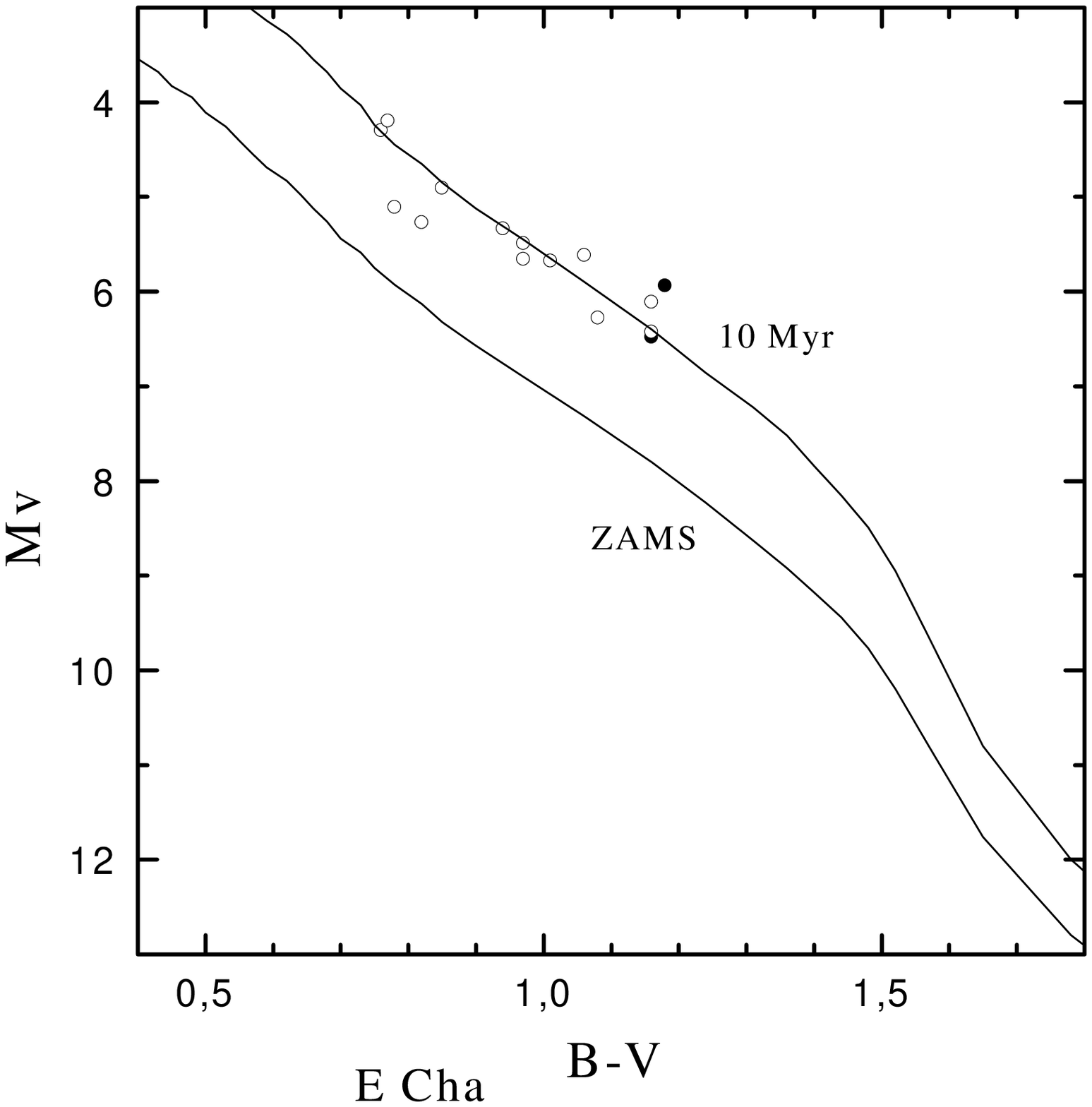,width=5.7cm}}
\caption{Evolutionary diagram for $\epsilon$ ChaA .}}
\end{figure}
\pagebreak
\begin{figure}[ht]
\centerline{\psfig{file=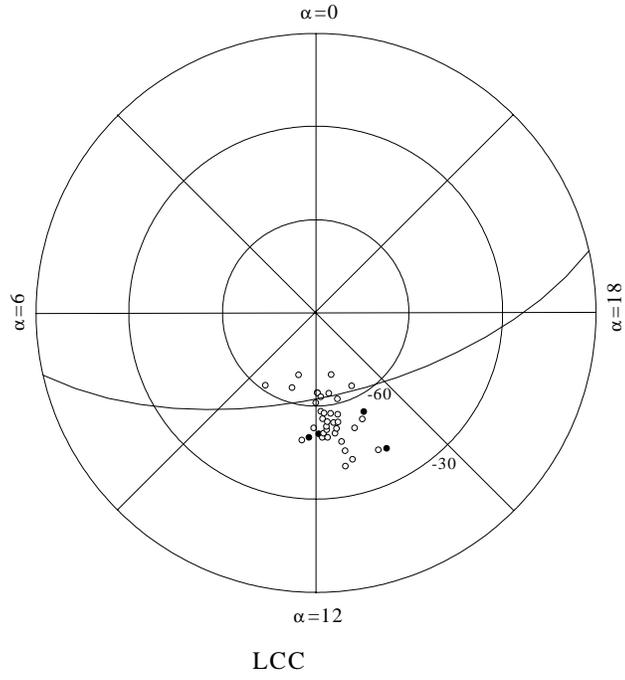,height=9cm}}
\caption{Polar representation of the LCC stars.}
\end{figure} 

\begin{figure}[ht]
\sidebyside
{{\psfig{file=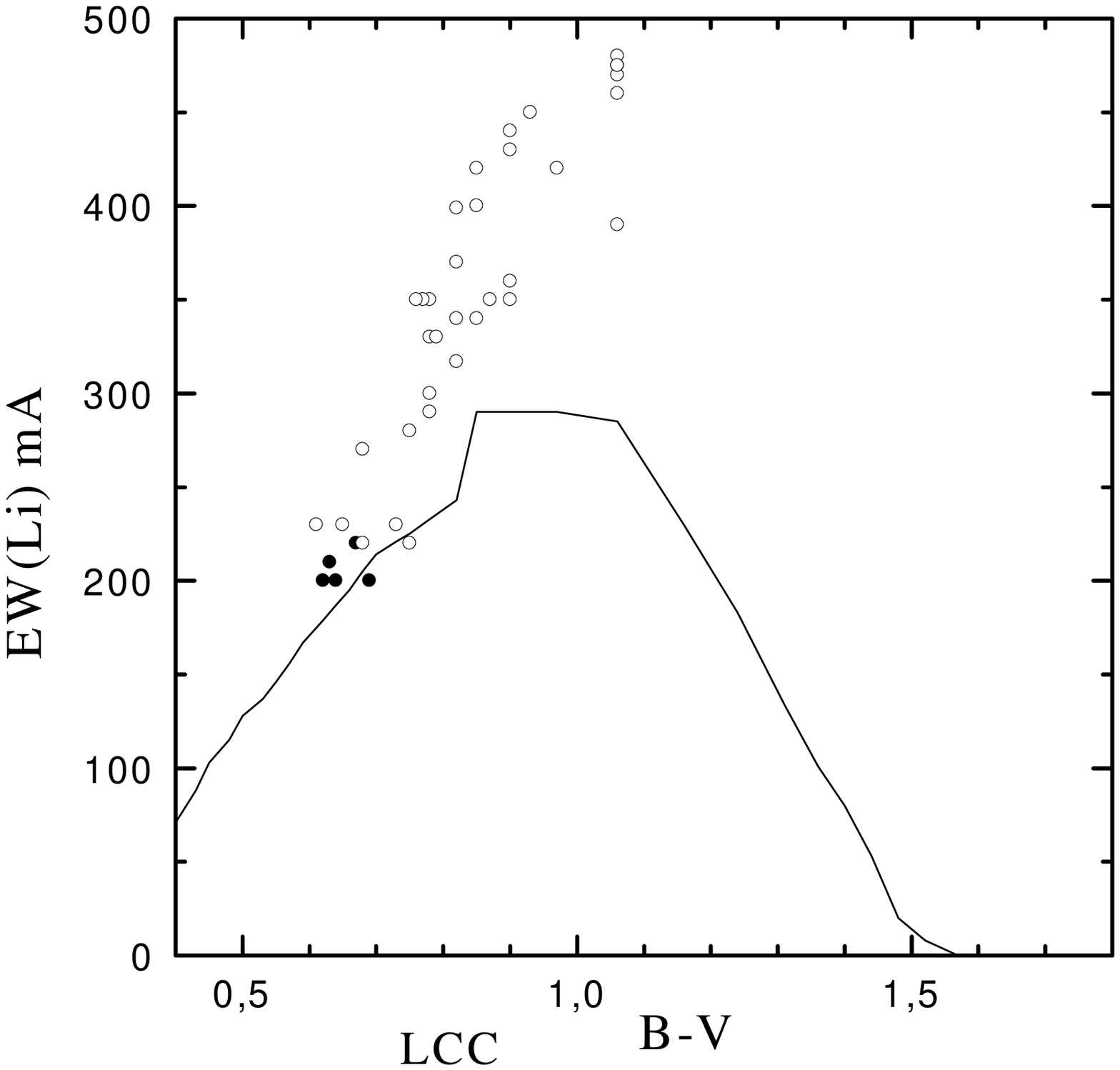,width=5.7cm}}
\caption{Distribution of Li line equivalent width for LCC.}}
{{\psfig{file=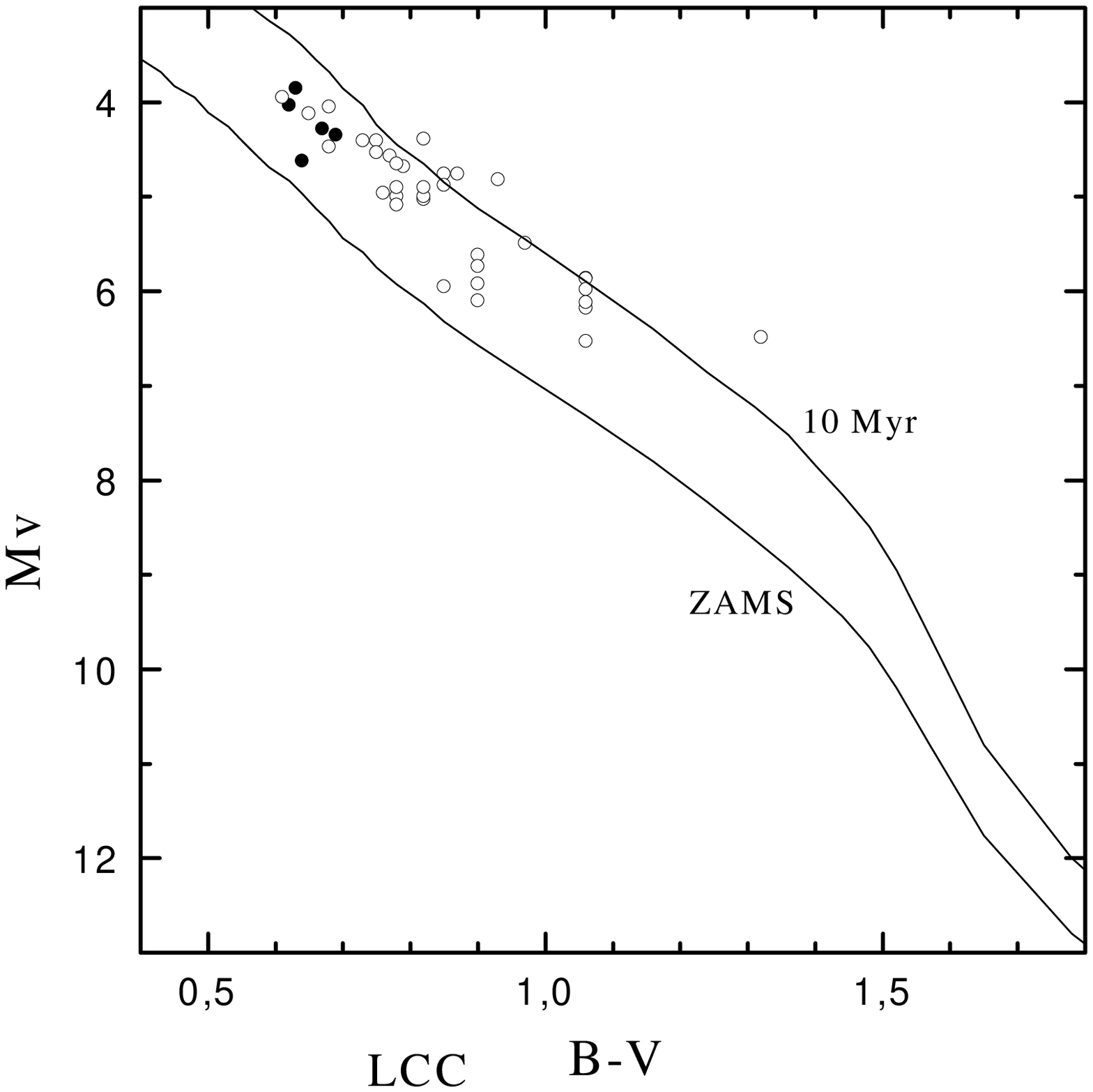,width=5.7cm}}
\caption{Evolutionary diagram for LCC.}}
\end{figure}
\pagebreak
\begin{figure}[ht]
\centerline{\psfig{file=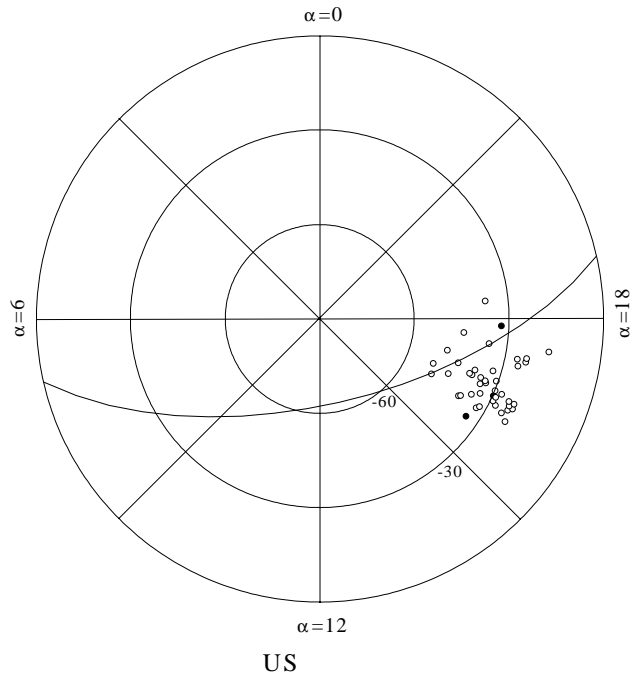,height=9cm}}
\caption{Polar representation of the US stars.}
\end{figure} 

\begin{figure}[ht]
\sidebyside
{{\psfig{file=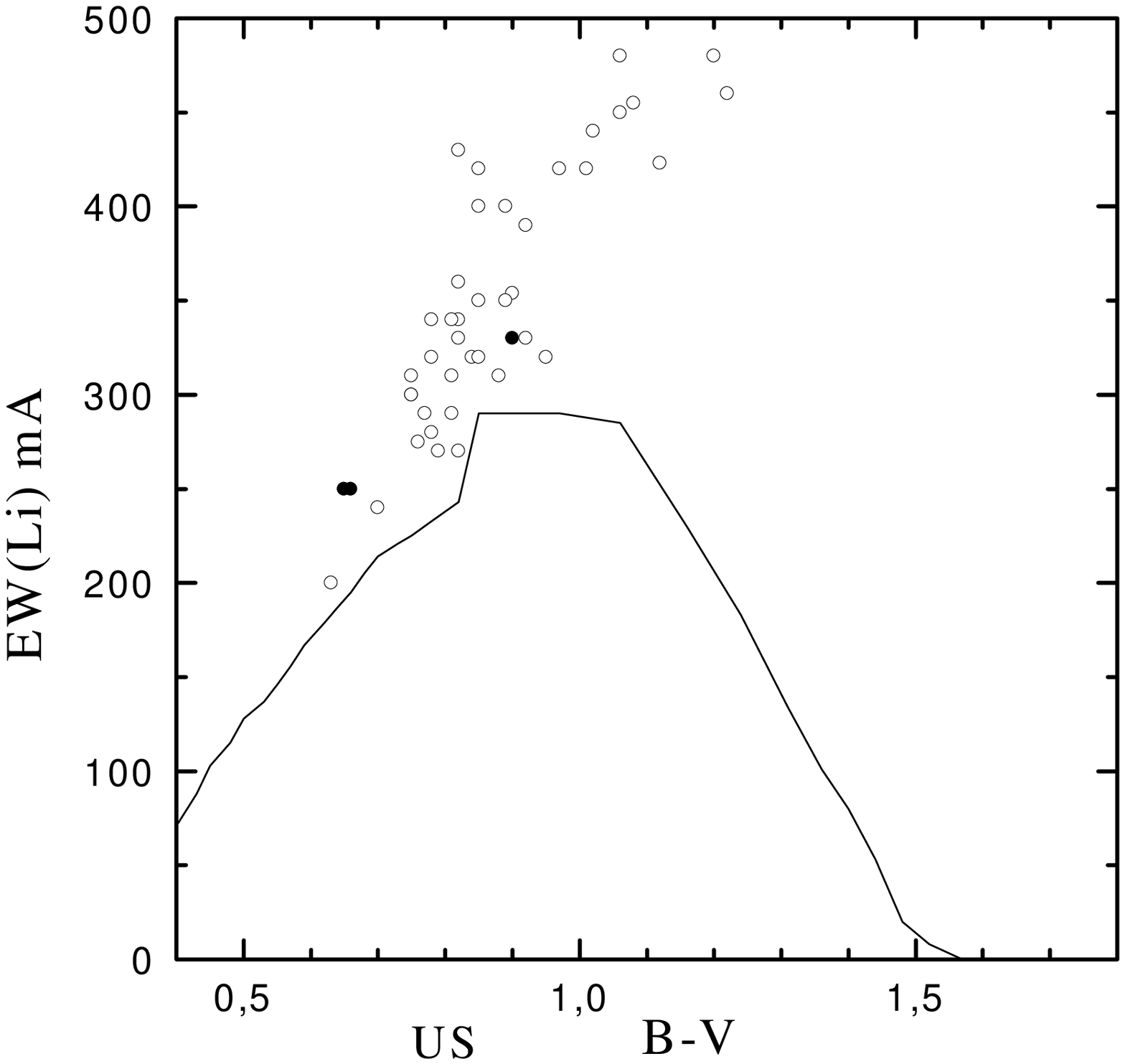,width=5.7cm}}
\caption{Distribution of Li line equivalent width for US.}}
{{\psfig{file=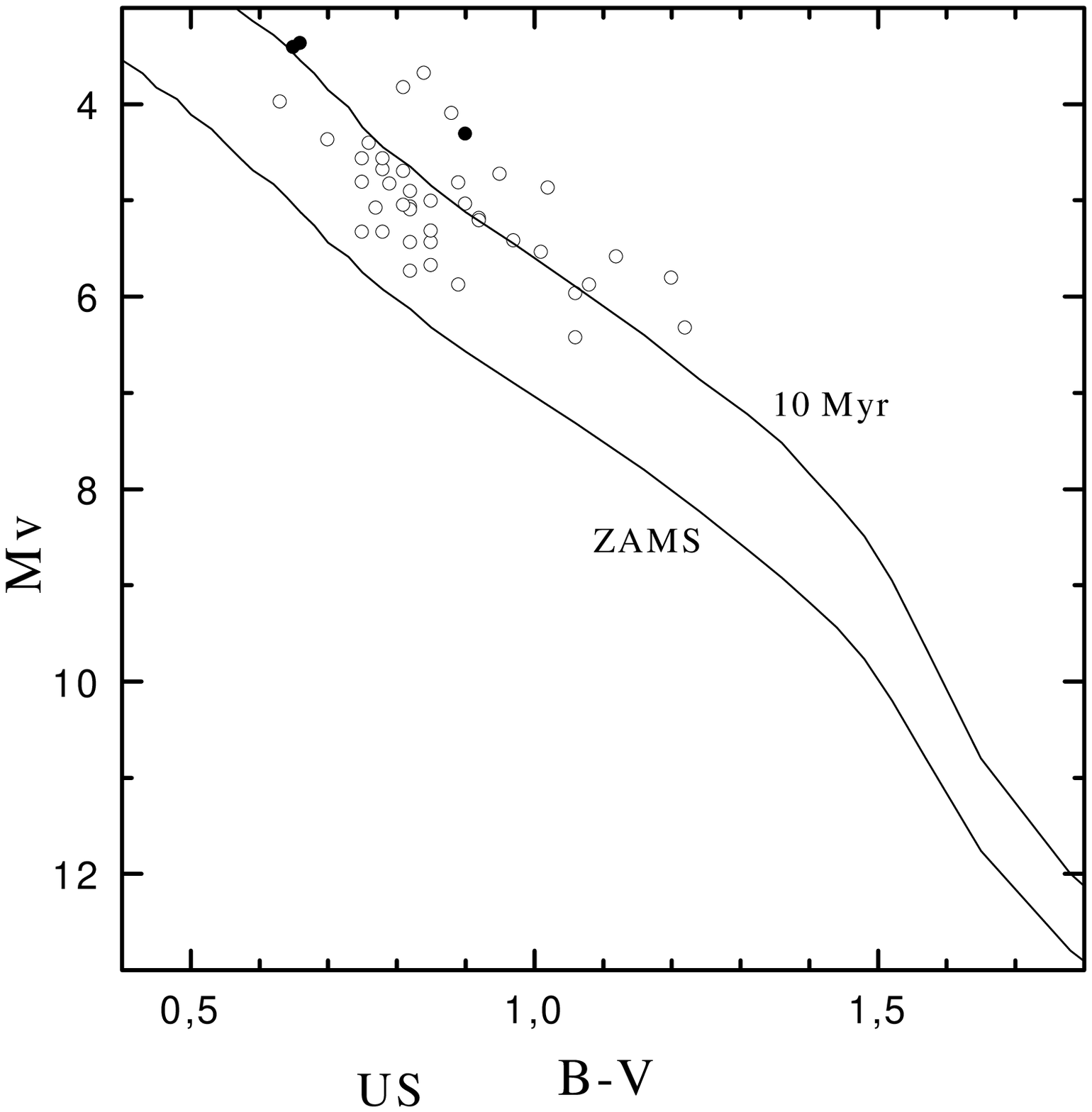,width=5.7cm}}
\caption{Evolutionary diagram for US.}}
\end{figure}
\pagebreak
\begin{figure}[ht]
\centerline{\psfig{file=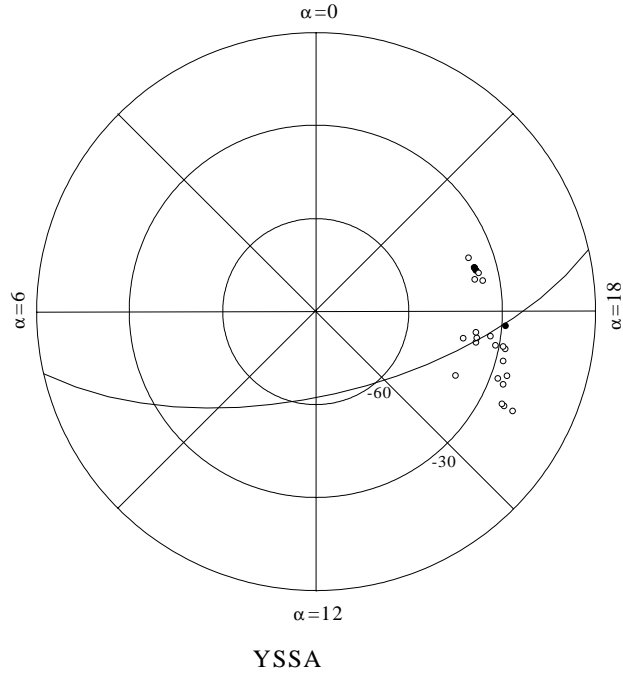,height=9cm}}
\caption{Polar representation of the YSSA stars.}
\end{figure} 

\begin{figure}[ht]
\sidebyside
{{\psfig{file=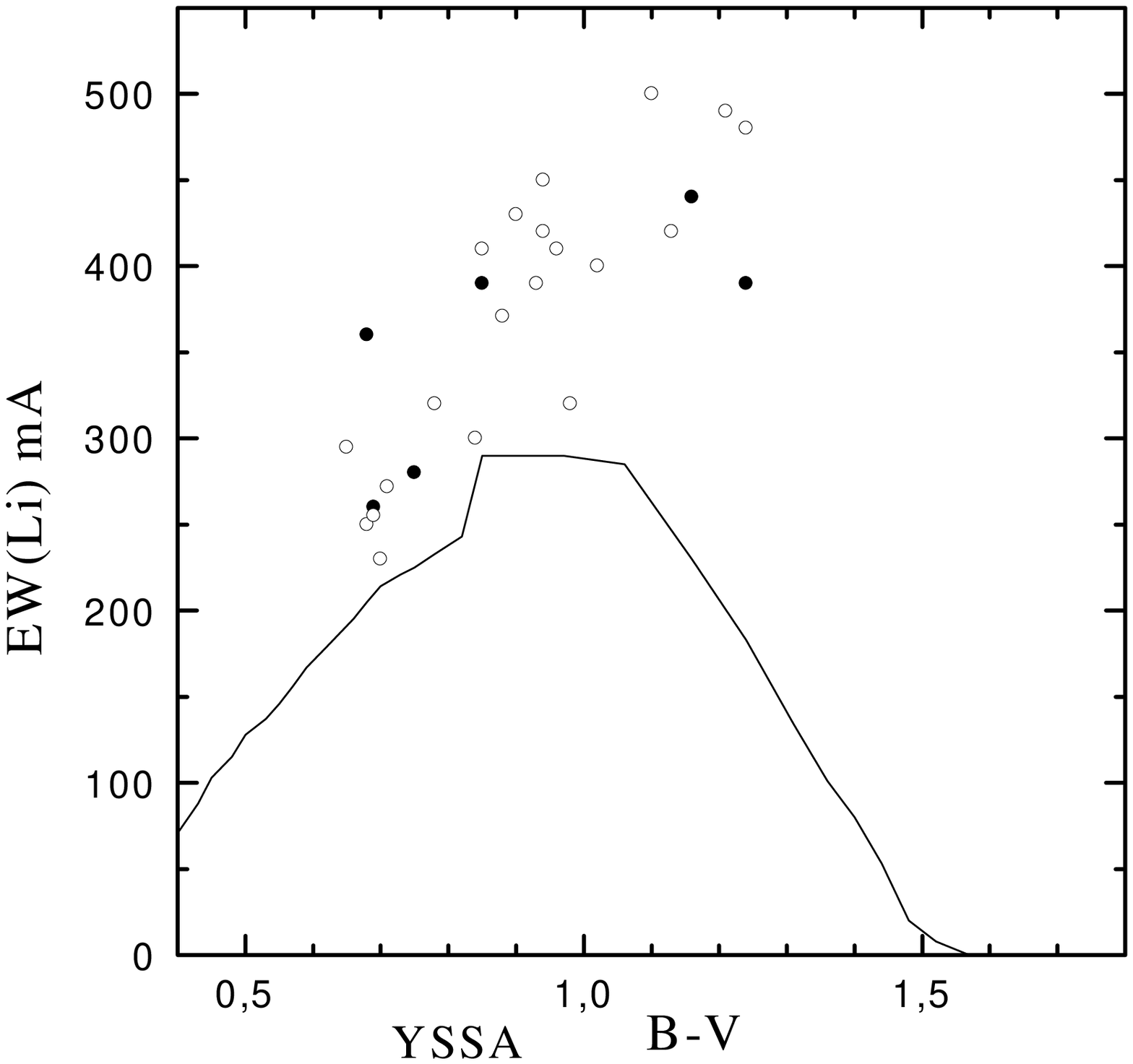,width=5.7cm}}
\caption{Distribution of Li line equivalent width for YSSA.}}
{{\psfig{file=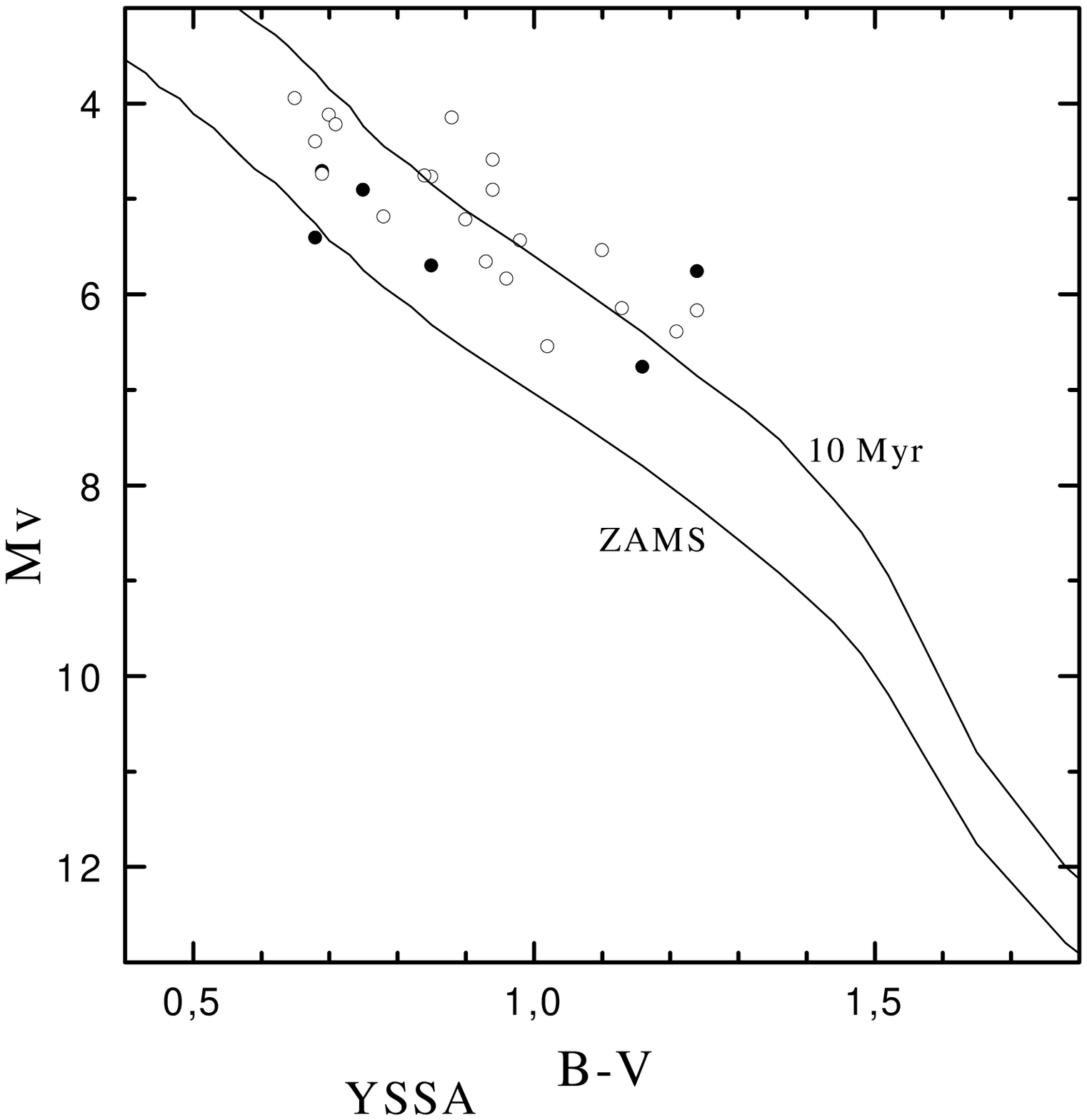,width=5.7cm}}
\caption{Evolutionary diagram for YSSA.}}
\end{figure}
\pagebreak
\begin{figure}[ht]
\centerline{\psfig{file=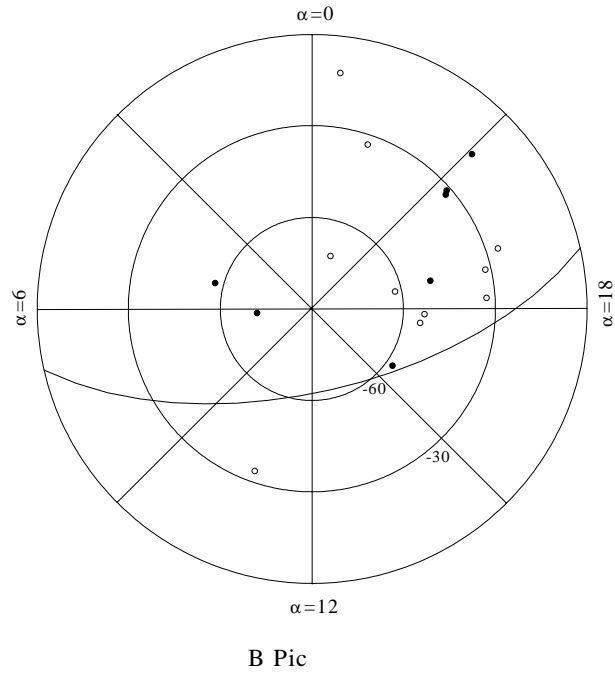,height=9cm}}
\caption{Polar representation of the  $\beta$ PicA stars.}
\end{figure} 

\begin{figure}[ht]
\sidebyside
{{\psfig{file=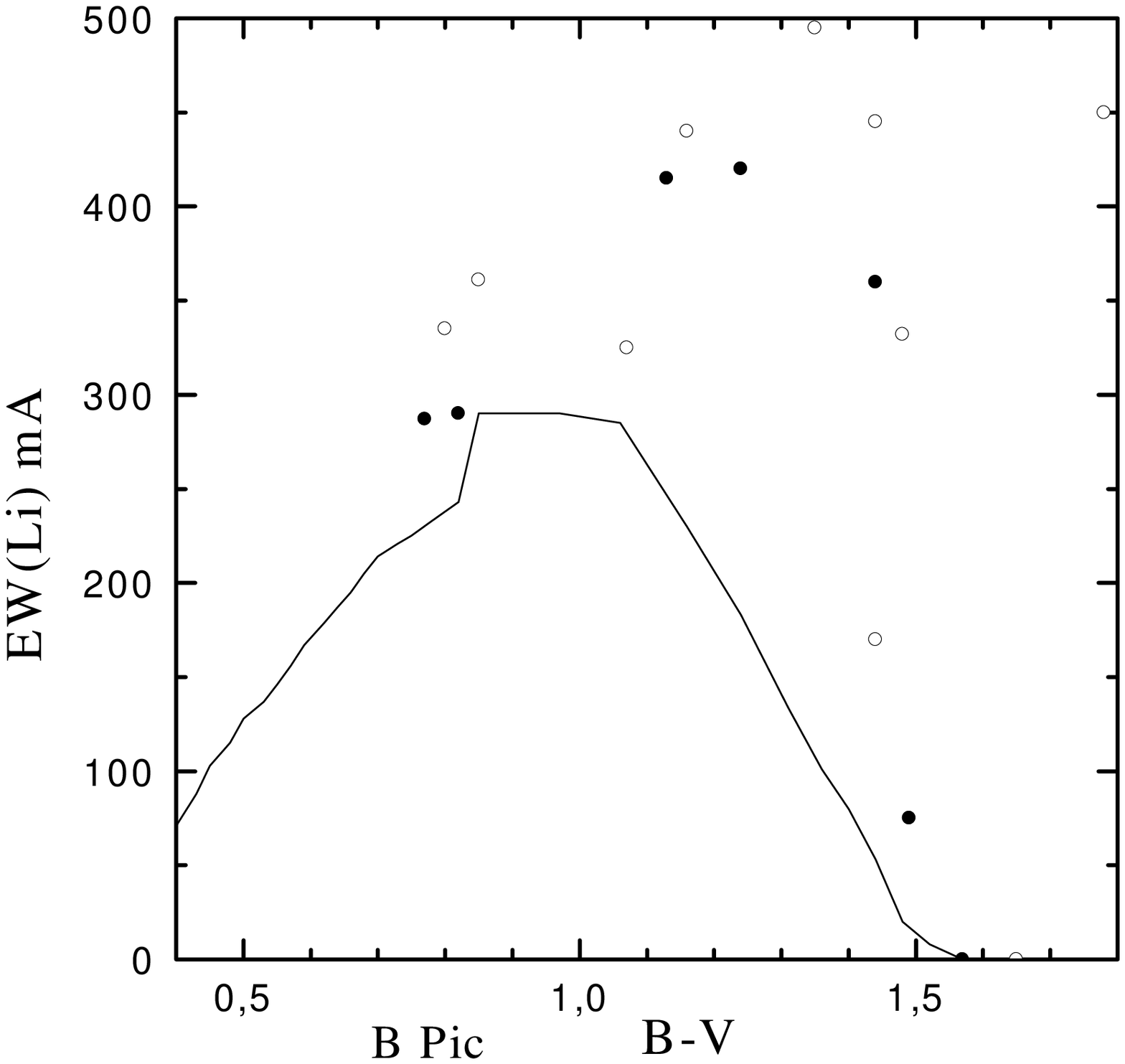,width=5.7cm}}
\caption{Distribution of Li line equivalent width for $\beta$ PicA.}}
{{\psfig{file=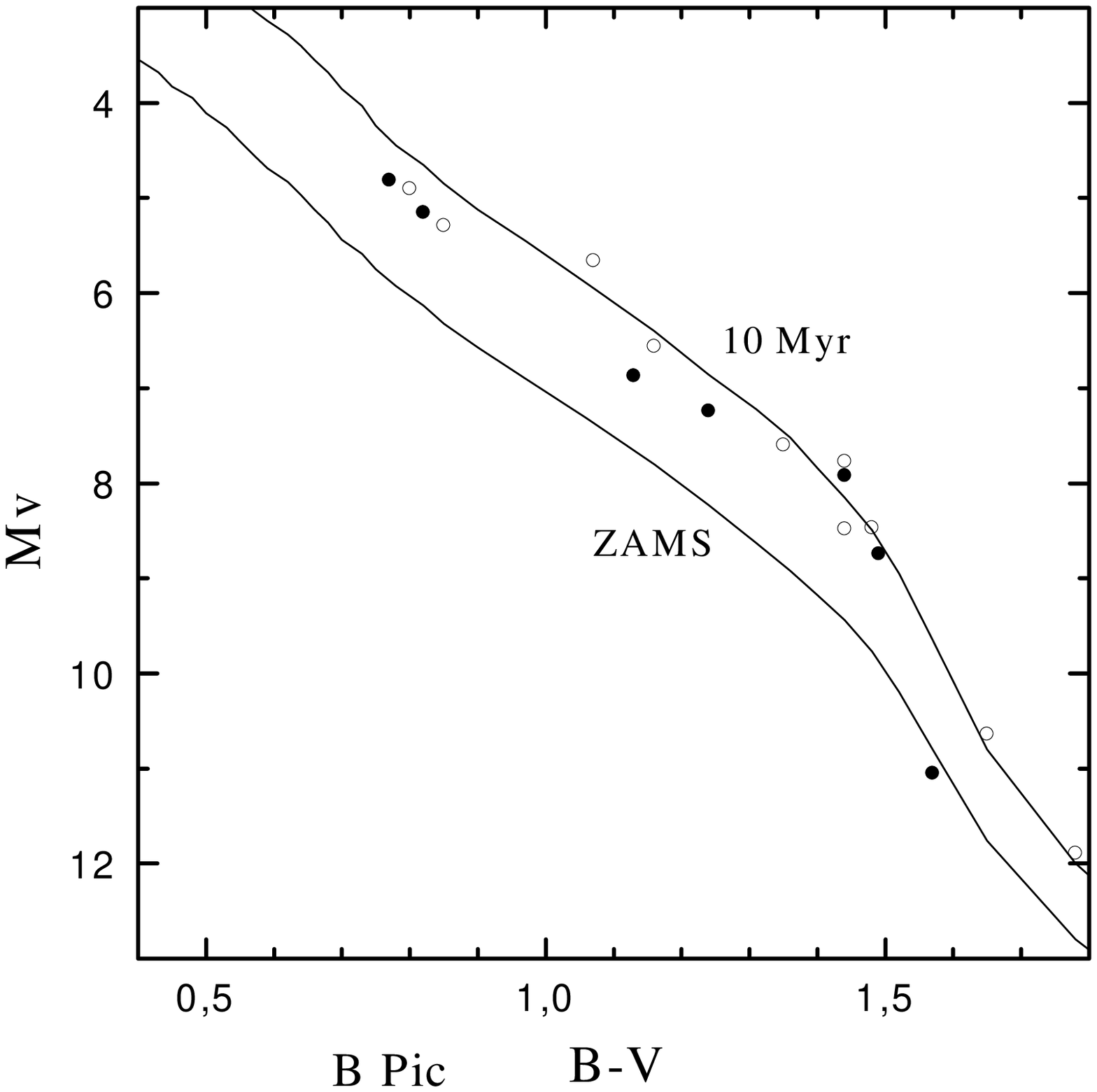,width=5.7cm}}
\caption{Evolutionary diagram for $\beta$ PicA.}}
\end{figure}
\pagebreak
\begin{figure}[ht]
\centerline{\psfig{file=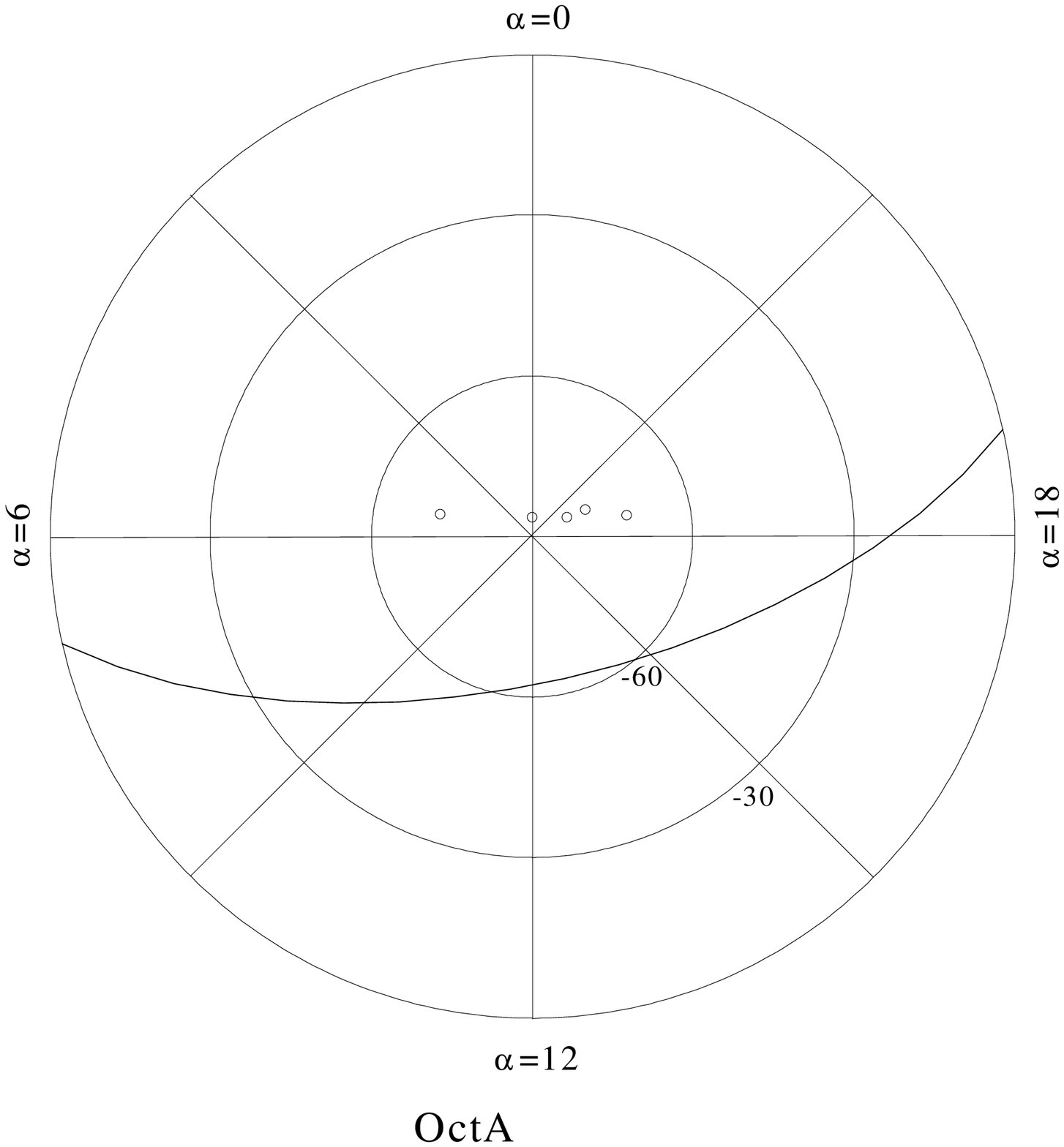,height=9cm}}
\caption{Polar representation of the OctA stars.}
\end{figure} 

\begin{figure}[ht]
\sidebyside
{{\psfig{file=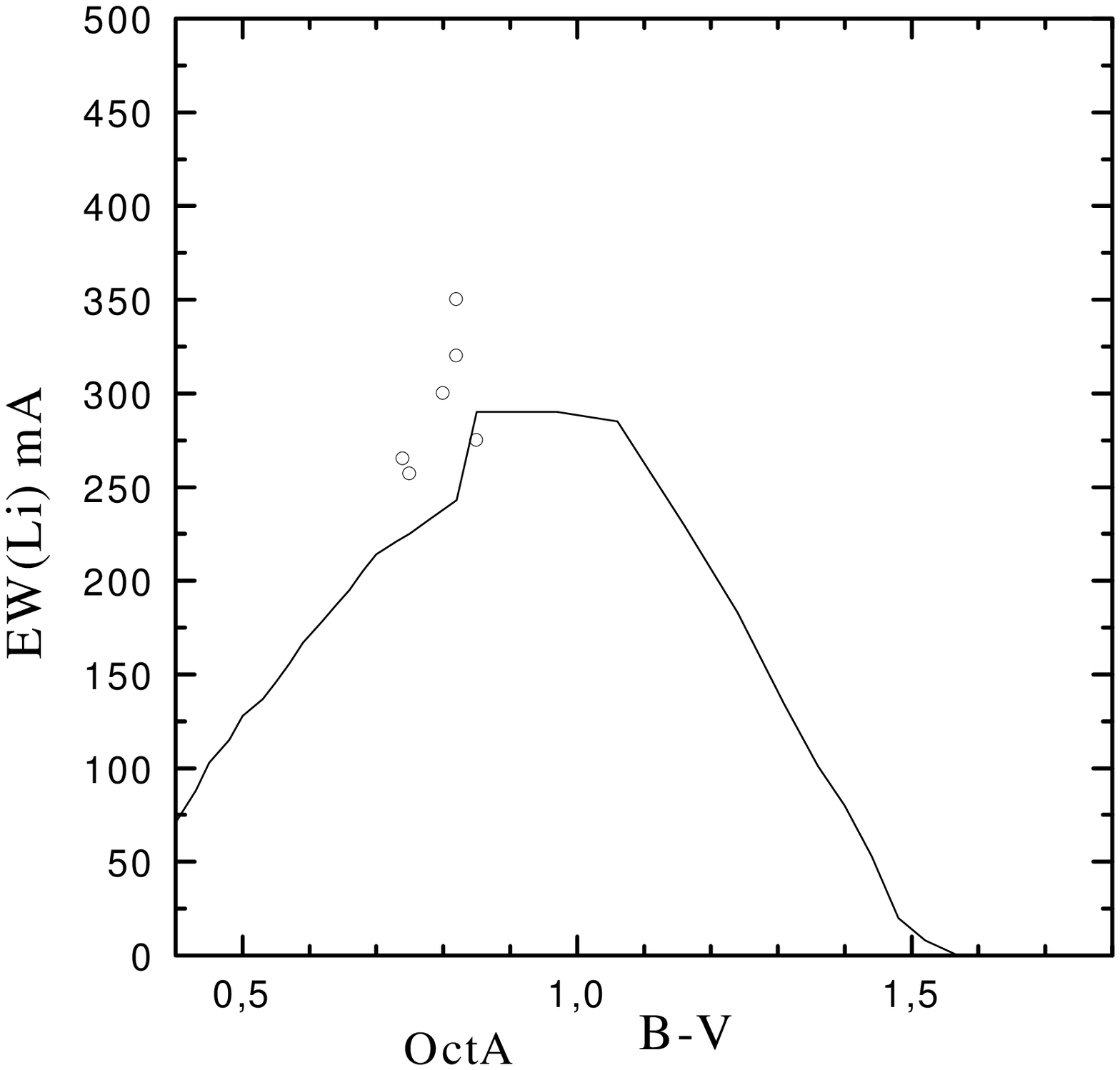,width=5.7cm}}
\caption{Distribution of Li line equivalent width for OctA.}}
{{\psfig{file=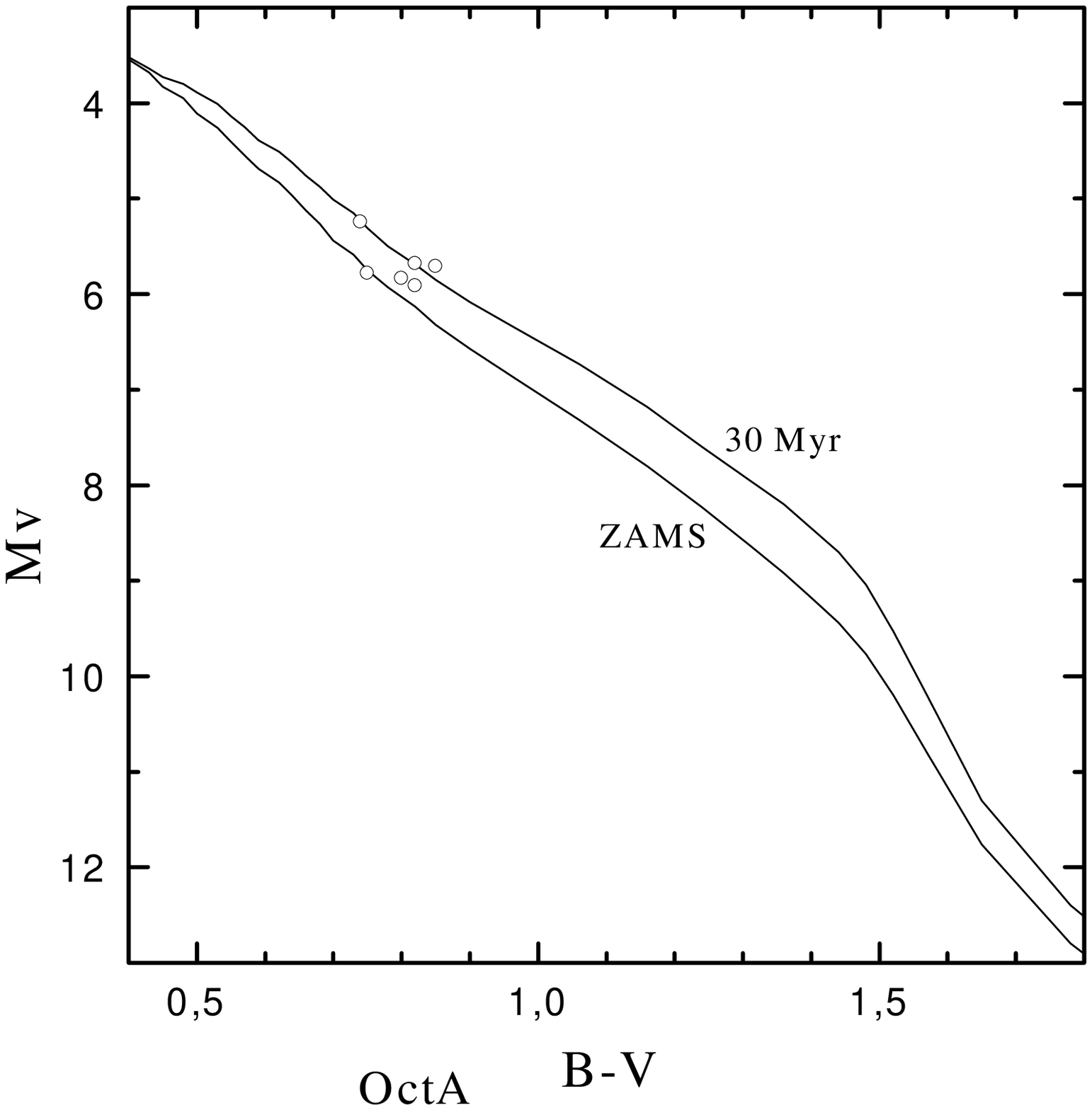,width=5.7cm}}
\caption{Evolutionary diagram for OctA.}}
\end{figure}
\pagebreak
\begin{figure}[ht]
\centerline{\psfig{file=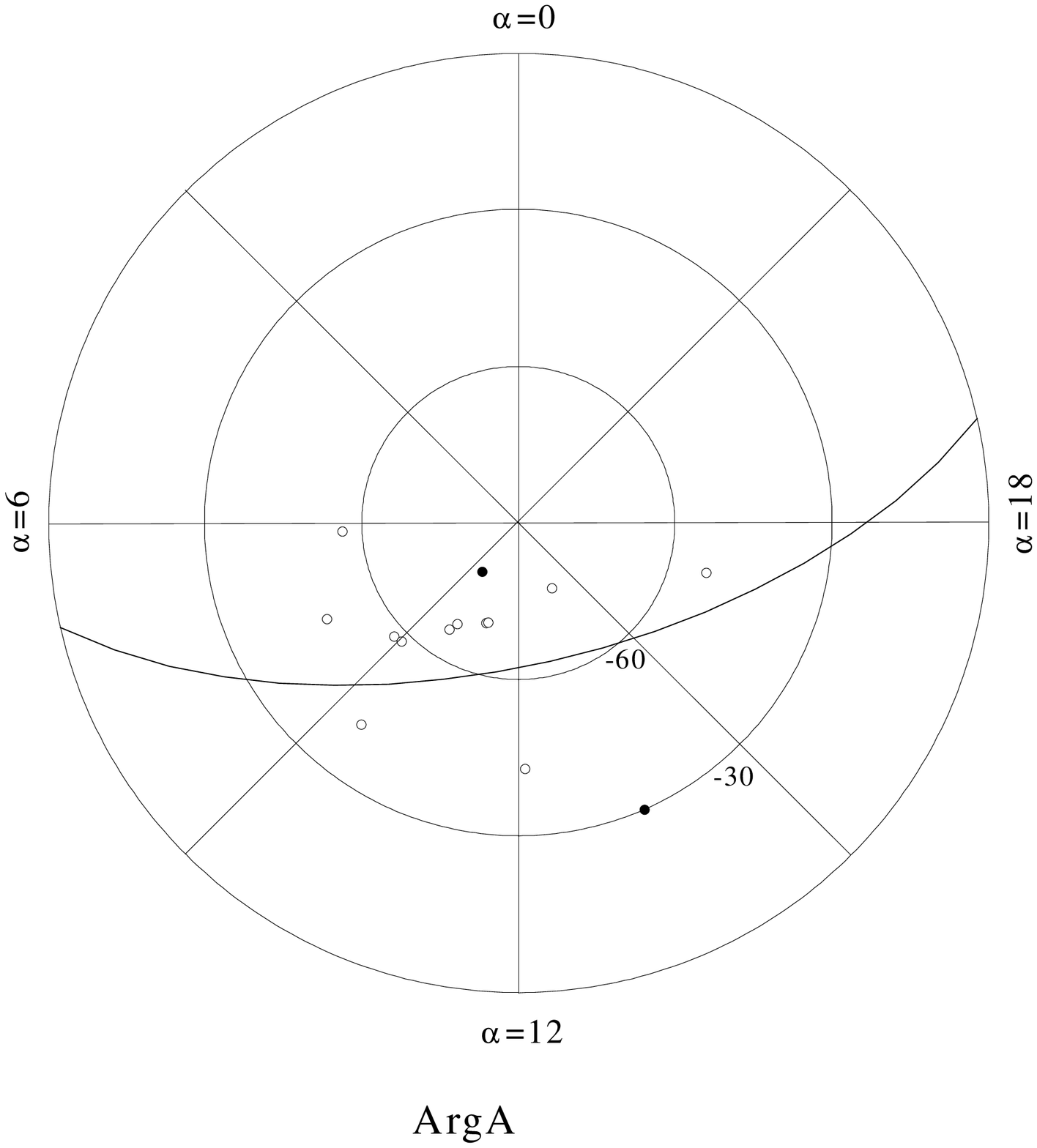,height=9cm}}
\caption{Polar representation of the ArgusA stars.}
\end{figure} 

\begin{figure}[ht]
\sidebyside
{{\psfig{file=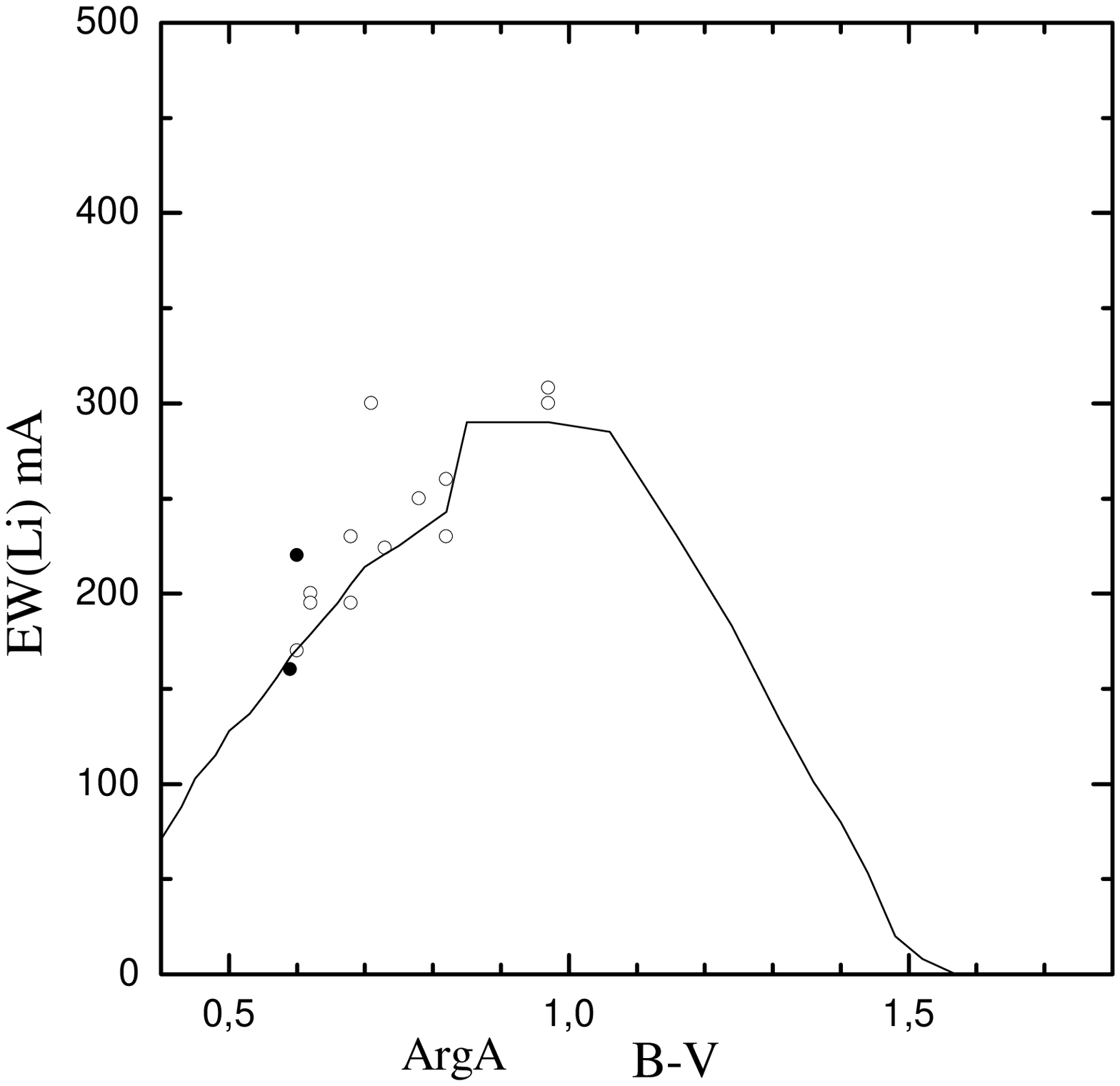,width=5.7cm}}
\caption{Distribution of Li line equivalent width for ArgusA.}}
{{\psfig{file=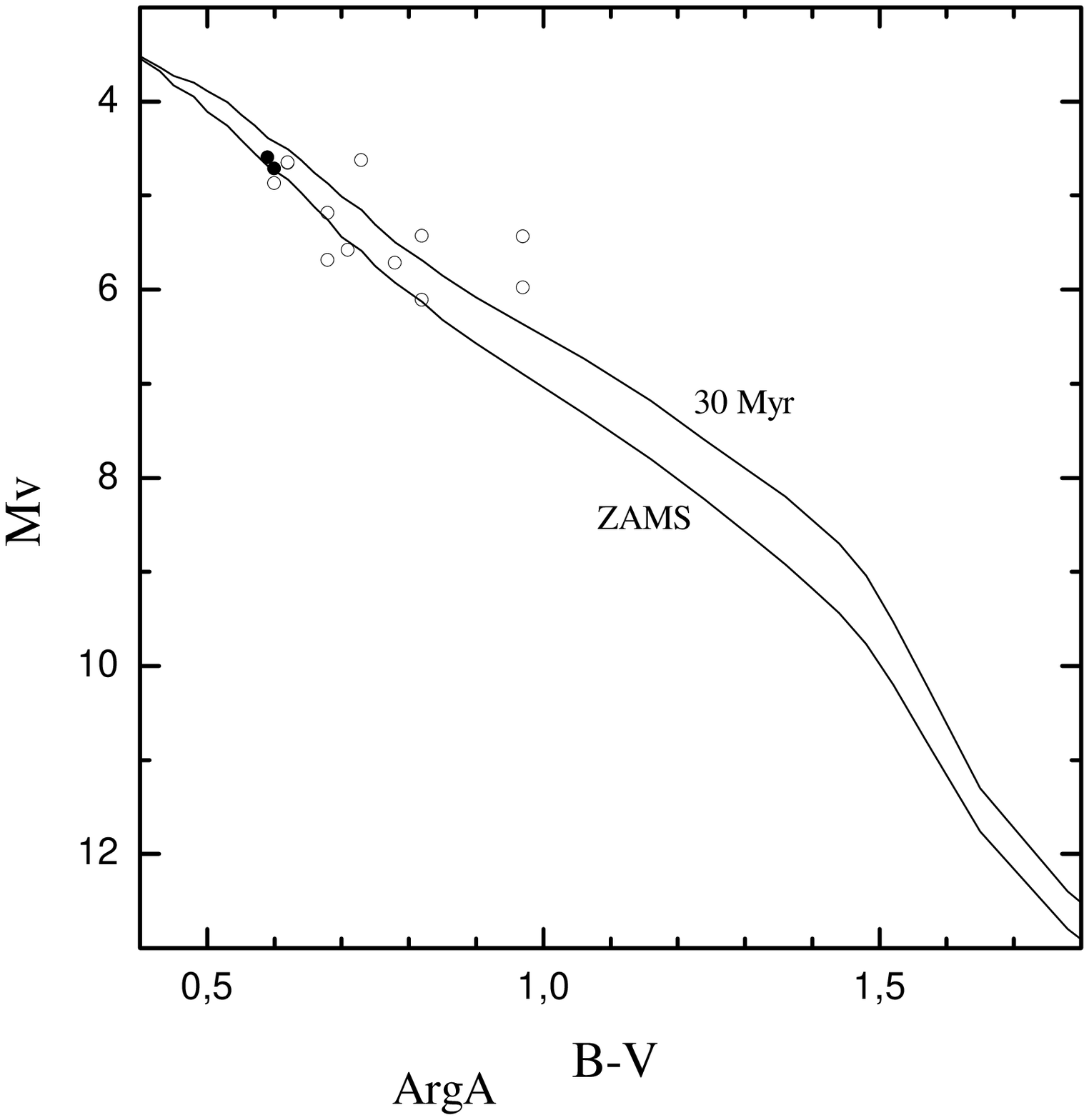,width=5.7cm}}
\caption{Evolutionary diagram for ArgusA.}}
\end{figure}
\pagebreak
\begin{figure}[ht]
\centerline{\psfig{file=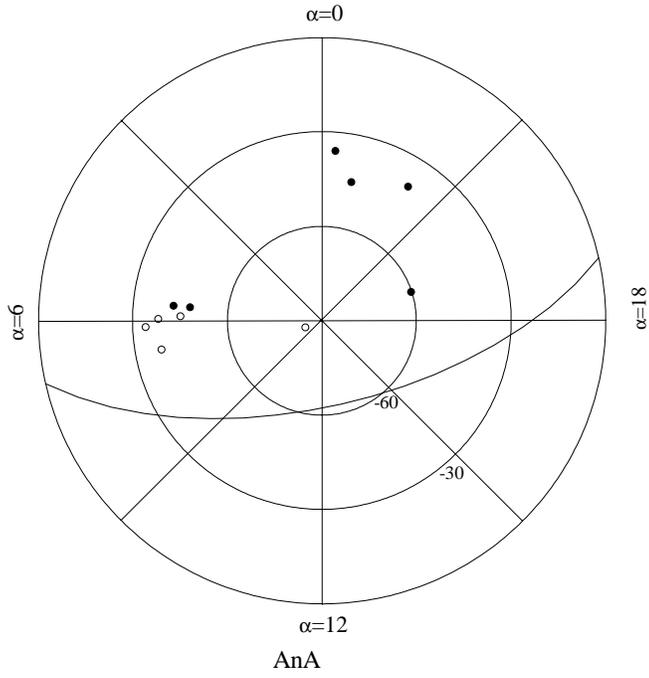,height=9cm}}
\caption{Polar representation of the AnA stars.}
\end{figure} 

\begin{figure}[ht]
\sidebyside
{{\psfig{file=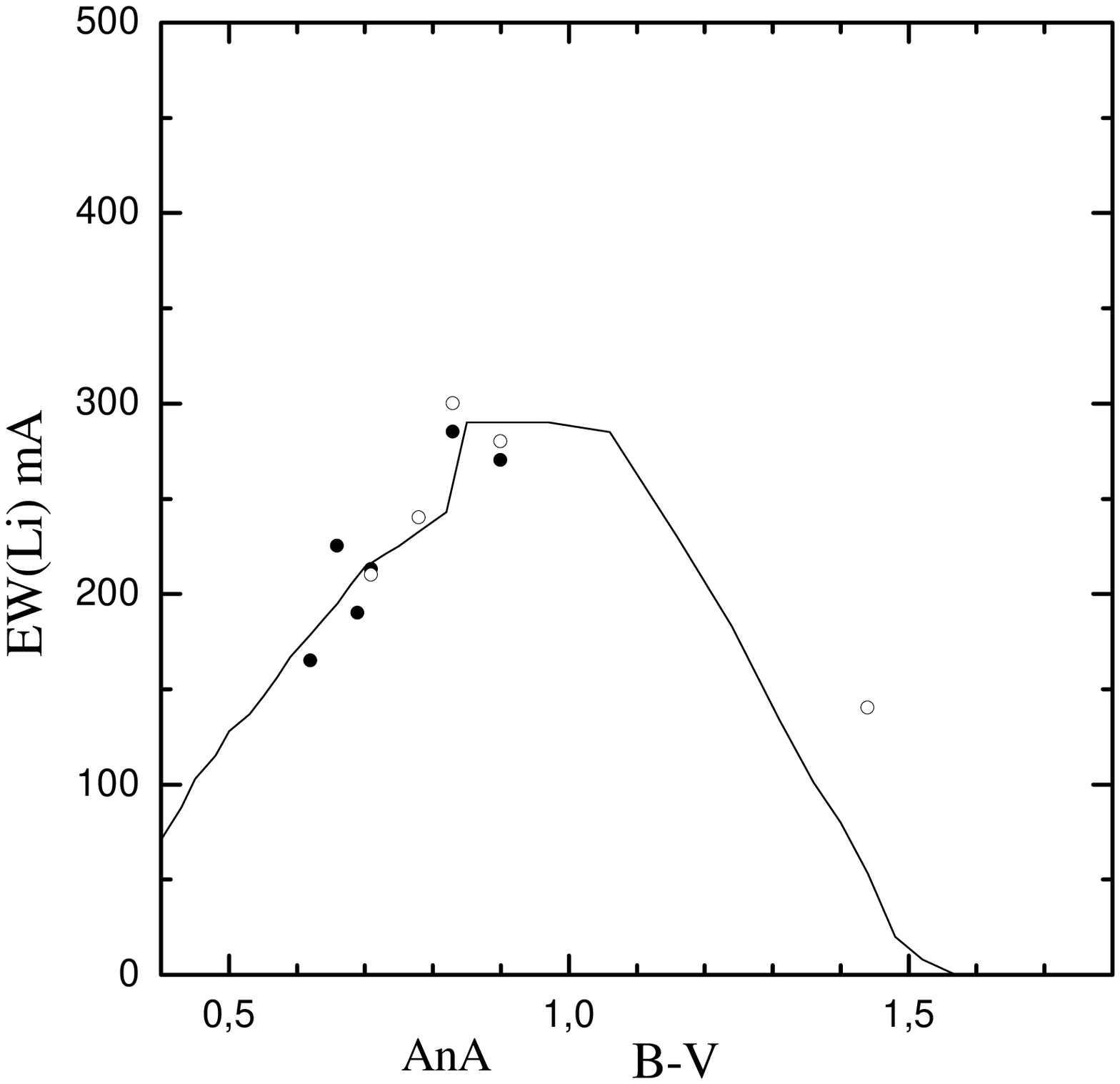,width=5.7cm}}
\caption{Distribution of Li line equivalent width for AnA.}}
{{\psfig{file=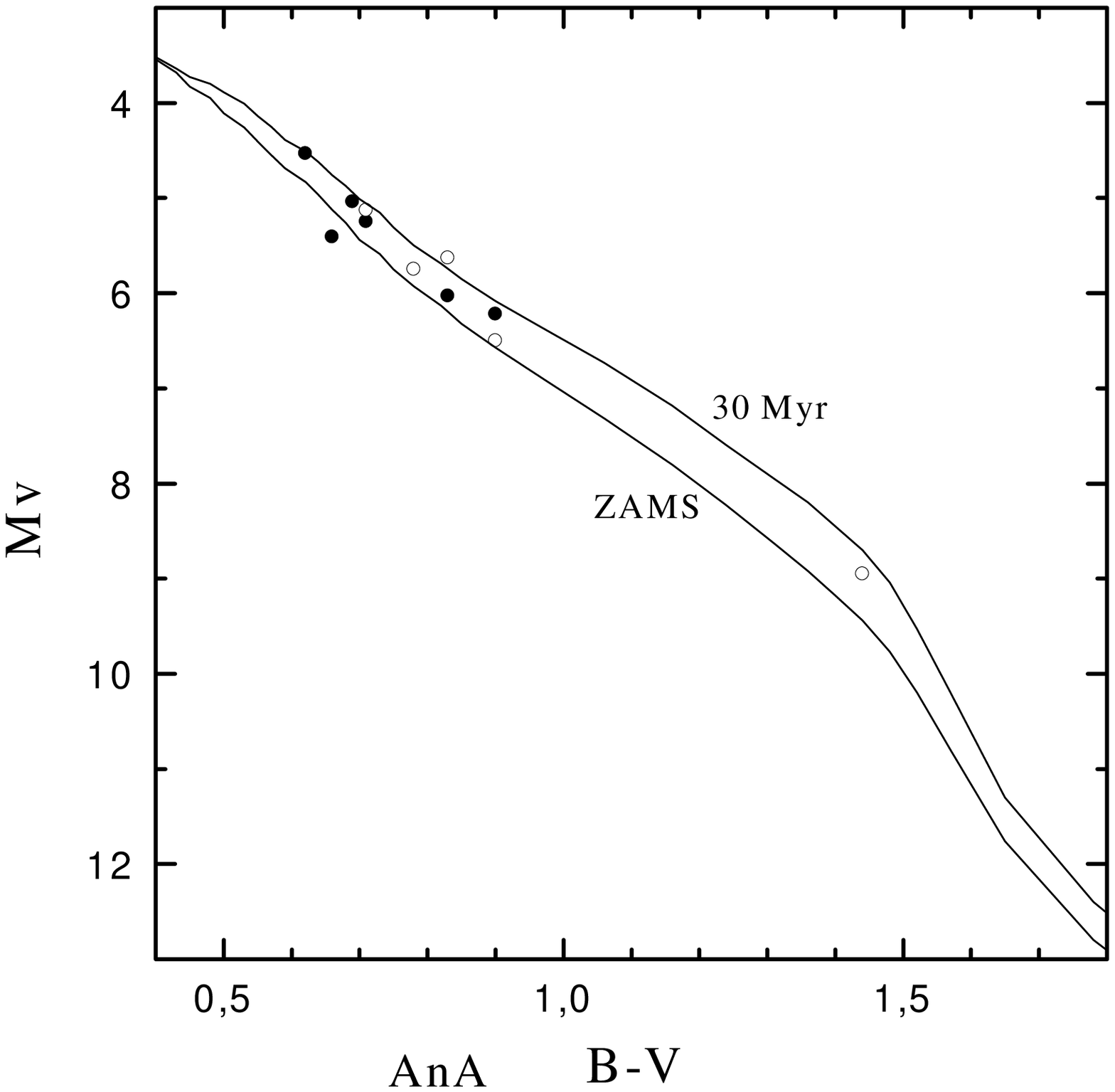,width=5.7cm}}
\caption{Evolutionary diagram for AnA.}}
\end{figure}



\begin{chapthebibliography}{1}

\bibitem{covino}
Covino, E., Alcal\'{a}, J.M., Allain, S., Bouvier, J., Terranegra, L.,
\& Krautter, J. 1997, A\&A, 328, 187
\bibitem{reza01}
de la Reza, J. R., da Silva, L., Jilinski, E., Torres, C. A. O., 
\& Quast, G. R. 2001, in ASP Conf. Ser. Vol. 244, Young stars near earth: 
progress and prospects, ed. R. Jayawardhana \& T. P. Greene (San Francisco: ASP), 37
\bibitem{reza86}
de la Reza, R., Quast, G. R., Torres, C. A. O., Mayor \& M. Vieira, G.V. 
1986, in Symp.NASA-ESA. New Insights in Astrophysics, ESA S-263, 107
\bibitem{reza89}
de la Reza, R., Torres, C. A. O., Quast, G. R., Castilho, B.V., 
\& Vieira, G.L.  1989, ApJL, 343, L61
\bibitem{hetem92}
Gregorio-Hetem, J., L\'{e}pine, J. R. D., Quast, G. R., Torres, C. A. O., 
\& de la Reza, R.  1992, AJ, 103, 549
\bibitem{jab94}
Jablonski, F., Baptista, R., Barroso, J., Jr., Gneiding, C. D., 
Rodrigues, F., \& Campos, R. P. 1994, PASP, 106, 1172

\bibitem{jef95}
Jeffries, R. D. 1995, MNRAS, 273, 559
\bibitem{kaufer}
Kaufer, A., Stahl, S., Tubbesing, S., Norregaard, P., Avila, G.,
Francois, P., Pasquini, L., \& Pizzella, A. 1999, Messenger, 95, 8
\bibitem{mama}
Mamajek, E. E., Lawson, W. A., \& Feigelson, E. D. 2000, ApJ, 544, 356 
\bibitem{neu97}
Neuh\"{a}user, R. 1997, Science, 276, 1363
\bibitem{neu00}
Neuh\"{a}user, R., Walter, F.  M., Covino, E.,
Alcal\'a, J.  M., Wolk, S.  J., Frink, S., Guillout, P.,
Sterzik, M.  F., \&  Comer\'on, F. 2000, A\&AS, 146, 323
\bibitem{quast98}
Quast, G.R. 1998, thesis ON-Rio de Janeiro
\bibitem{quast01}
Quast, G. R., Torres, C. A. O., de la Reza, J. R., da Silva, L., \&
Drake, N. 2001, in ASP Conf. Ser. Vol. 244, Young stars near earth: 
progress and prospects, ed. R. Jayawardhana \& T. P. Greene (San Francisco: ASP), 49
\bibitem{quast00}
Quast, G.R.; Torres, C. A. O.; de la Reza, R., da Silva, L., \& Mayor, M. 
2000, IAU Symposium No. 200 ``Birth and Evolution of Binary Stars'', Potsdam, Germany, 28
\bibitem{queloz}
Queloz, D., Mayor, M., Naef, D., Santos, N., Udry, S., Burnet,
M., \& Confino, B. 2000, in VLT Opening Symposium
From Extrasolar Planets to Brown Dwarfs, ESO Astrophys. Symp., Springer
Verlag, Heidelberg, 548
\bibitem{sartori}
Sartori, M. J., L\'epine, J. R. D., \& Dias, W. S. 2003, A\&A, 404, 913
\bibitem{siess} 
Siess, L., Forestini, M., \& Dougados, C. 1997, A\&A, 324, 556
\bibitem{song02}
Song, I., Bessel, M. S., \& Zuckerman, B. 2002 ApJL, 581, L434
\bibitem{torres95}
Torres, C. A. O., Quast, G., de la Reza, R.,  Gregorio-Hetem, J., \& L\'{e}pine, J. R. D.  
1995, AJ, 109, 2146
\bibitem{torres98}
Torres, C. A. O. 1998, Publica\c{c}\~{a}o Especial do Observat\'{o}rio Nacional, 10/99
\bibitem{torres00}
Torres, C. A. O., da Silva, L.,  Quast, G., de la Reza, R., \&
Jilinski, E.  2000, AJ, 120, 1410
\bibitem[Torres et al., 2001]{torres01}
Torres, C. A. O., Quast, G. R., de la Reza, J. R., da Silva, L., 
Melo, \& C. H. F. 2001, in ASP Conf. Ser. Vol. 244, Young stars near earth: 
progress and prospects, ed. R. Jayawardhana \& T. P. Greene (San Francisco: ASP), 43
\bibitem{gt03}
Torres, G., Guenther, E. W., Marschall, L. A., Neuh\"{a}user, R., Latham, D. W., 
\& Stefanik, R. P. 2003, AJ 125, 825 
\bibitem{zuc01}
Zuckerman, B.,  Sing, I.,  Bessell, M. S., \& Webb, R. A  2001,  ApJL, 562, L87
\bibitem{zuc00}
Zuckerman, B., \& Webb, R. A. 2000, ApJ, 535, 959


\end{chapthebibliography}
\end{document}